\documentclass{easees2013}
\usepackage{graphicx}
\usepackage[authoryear]{natbib}
\usepackage[fleqn]{amsmath}
\usepackage[]{amsfonts}%
\usepackage{xcolor}
\usepackage{xspace}
\usepackage{bm}
\usepackage{amssymb}
\usepackage[hyphens]{url}
\usepackage[colorlinks=true,linkcolor=red,urlcolor=blue,citecolor=blue]{hyperref}
\usepackage{breakurl}

\newcommand{\btheta}{\vec{\bm \vartheta}}
\newcommand{\bdata}{\vec{\mathbf D}}
\newcommand{\like}{\mathcal{L}}

\def\Msun{M_{\odot}}
\def\Nstar{N_{\star}}
\def\boxes{\prod_{i=1}^B}
\def\zapb{\boxes{(m_i+s_i)!\over m_i!s_i!}}

\def\beq{\begin{equation}}
\def\eeq{\end{equation}}
\def\ltsima{$\; \buildrel < \over \sim \;$}
\def\simlt{\lower.5ex\hbox{\ltsima}}
\def\gtsima{$\; \buildrel > \over \sim \;$}
\def\simgt{\lower.5ex\hbox{\gtsima}}

\def\yr{{\rm\,a}\xspace}

\newcommand\ion[2]{#1$\;${\small\rmfamily\@Roman{#2}}\relax}%

\def\s{\ifmmode \widetilde \else \~\fi}
\def\={\overline}

\def\spose#1{\hbox to 0pt{#1\hss}}

\def\cf{{\it cf.\xspace}}
\def\lta{\mathrel{\spose{\lower 3pt\hbox{$\mathchar"218$}}
     \raise 2.0pt\hbox{$\mathchar"13C$}}}
\def\gta{\mathrel{\spose{\lower 3pt\hbox{$\mathchar"218$}}
     \raise 2.0pt\hbox{$\mathchar"13E$}}}
\def\Dt{\spose{\raise 1.5ex\hbox{\hskip3pt$\mathchar"201$}}}    
\def\dt{\spose{\raise 1.0ex\hbox{\hskip2pt$\mathchar"201$}}}    

\def\dotsfill{\leaders\hbox to 1em{\hss.\hss}\hfill}

\newcommand{\an}{Ast. Nach.}
\newcommand{\aj}{AJ}
\newcommand{\apj}{ApJ}
\newcommand{\apjs}{ApJS}
\newcommand{\apjl}{ApJ Letters}
\newcommand{\aap}{A\&A}
\newcommand{\aaps}{A\&AS}
\newcommand{\mnras}{MNRAS}
\newcommand{\advast}{Adv. Ast.}

\newcommand{\araa}{ARA\&A}

\newcommand{\memsai}{Mem. Soc. Ast. Ital.}
 
\newcommand{\apss}{Ap\&SS}

\newcommand{\pasp}{Pub. Ast. Soc. Pacific}

\newcommand{\pasa}{Pub. Ast. Soc. Aus.}

\newcommand{\rmp}{Rev. Mod. Physics}

\newcommand{\newast}{New Ast.}
\newcommand{\rpp}{Rep. Prog. Phys.}
\newcommand{\aar}{Ast. Astrophys. Rev.}
\newcommand{\jbaa}{Jour. Brit. Ast. Assoc.}

\DeclareGraphicsExtensions{.pdf}
\TitreGlobal{The Ages of the Stars}
\begin{document}
\title{BAYESIAN ISOCHRONE FITTING \\[0.3cm]
AND STELLAR AGES}
\runningtitle{D. Valls-Gabaud: Bayesian Isochrone Fitting}
\author{David Valls-Gabaud}
\address{
 LERMA, CNRS UMR 8112, 
 Observatoire de Paris, 
 61 Avenue de l'Observatoire, 
 75014 Paris, France. Email \url{david.valls-gabaud@obspm.fr}}%
\begin{abstract}
Stellar evolution theory has been extraordinarily successful at explaining
the different phases under which stars form, evolve and die. While the
strongest constraints have traditionally come from binary stars, the advent
of asteroseismology is bringing unique measures in well-characterised stars.
For stellar populations in general, however, only photometric measures are
usually available, and the comparison with the predictions of stellar evolution
theory have mostly been qualitative. For instance, the geometrical shapes of
isochrones have been used to infer ages of coeval populations, but without
any proper statistical basis.  
In this chapter we provide a pedagogical review on a Bayesian
formalism to make quantitative inferences on the properties of single, binary
and small ensembles of stars, including unresolved populations. As an
example, we show
how stellar evolution theory can be used in a rigorous way as a prior 
information to measure the ages of stars between the ZAMS and the Helium 
flash, and their uncertainties, using photometric data only.
\end{abstract}
\maketitle
\section{Introduction}
\label{dvg:sec1}
In this chapter 
a brief summary is presented of the uses of stellar
evolution theory to infer properties of single stars 
(Section~\ref{dvg:single}), of detached binary stars whose components are assumed to
have evolved independently of each other (Section~\ref{dvg:double}), and coeval stellar populations
such as (presumably) those in clusters (Section~\ref{dvg:coeval}). 
When the fundamental properties of the star(s) in question are known 
(mass, absolute luminosity, effective temperature, etc), stellar tracks
computed for this particular (set of) star(s) can be used to infer further
properties, such as ages. In general, however, one wishes to use the
predictions of stellar evolution to infer these properties. The data at hand
are usually magnitudes and colours, hence the interpretation of the features
in the colour-magnitude diagrammes (CMDs) is carried out with isochrones rather
than stellar tracks. 
The thorny issue of transforming isochrones from/to the theoretical 
diagram to/from
 observed CMDs will not be dealt with here 
(see the contributions by Cassisi, and by Lebreton, Goupil and Montalb\'an 
in this volume), and constitute one of the sources
of systematic uncertainties. One also has to bear in mind
that while many efforts have been placed to
find the best transformations, the differences observed cannot (yet?) be 
fully ascribed
to either systematics in the observations or in missing/wrong physics in 
the stellar
evolutionary calculations. For instance the recent analysis by \cite{vandenberg2010} for some
globular clusters, and by \cite{an2007} for open clusters 
show that while some CMDs can be well fitted,
other colour-magnitude combinations of the \textsl{same} clusters show anomalies which go well
beyond the corrections for systematics, rotation, activity, transformations, metallicity, etc.
Empirical bolometric corrections are another source of 
uncertainty \citep{torres2010} as are 
the systematics in the determination of effective temperatures 
\citep[\eg,][]{ramirezmelendez2005}. Likewise, it does make sense to adopt a 
standard set of values \citep{harmanec2011} 
with nominal values to avoid some of the systematics arising with the 
adoption of different key
values (solar radius, mass, etc).
 Section~\ref{dvg:composite} deals
with the general problem of inverting the CMDs of a mixture of resolved
 stellar populations to
infer their distribution of ages and hence their
 chemical and star formation rate histories, while 
Section~\ref{dvg:pixels} is a very brief discussion on CMDs of pixels 
in unresolved 
stellar populations. Section~\ref{dvg:stats} closes this
chapter with a discussion on some 
statistical issues in the interpretation of CMDs. We will limit the scope of
this chapter to stars from the main sequence to the Helium flash, as the
predictions in this range appear to be the most robust ones. The white dwarf
phase can also be used in a rather robust way (provided the cooling and
the equation of state are properly characterised), as described by T. von Hippel
in this volume.

\section{Colour-Magnitude and Hertzsprung-Russell Diagrammes}
\label{dvg:sec2}
$ \; $ 
\vskip 2mm

\begin{flushright}
\begin{minipage}[b]{11.65cm}
\begin{quote}
\textsl{The plotting of the colors (or spectra) of stars as abscissae against their
absolute magnitudes (total magnitudes) has become one of the most lucrative
adventures in the study of star light.}
\end{quote}
\end{minipage}

 \cite{shapley1960} 
\end{flushright}

\vskip 3mm

It is appropriate to recall, in the context of this volume, that just over a century ago the 
first colour-magnitude diagram (CMD) was published. The
author of this landmark paper was not Ejnar Hertzsprung nor Henry N. Russell, but
 Hans O. Rosenberg, a colleague of Karl Schwarzschild at G\"ottingen. Rosenberg had been
working since 1907 on getting spectral properties of stars by measuring plates
 obtained with the Zeiss objective prism camera \citep{ejnar}. To maximise the number
of spectra per plate, he observed the Pleiades cluster and obtained spectra
for about 60 of them, over 1907--1909, noting that their inferred effective 
temperatures correlated with their apparent 
magnitudes in the first ever published CMD \citep{rosenberg1910}\footnote{A translation
into English is available at Leos Ondra's 
website \url{www.leosondra.cz/en/first-hr-diagram}}. His goal
was to ``\textsl{make the most accurate determination of the spectral types
of stars in the Pleiades}'' by using a ``\textsl{physiological blend}''
of the depth and width of the Ca~\textsc{ii} K line (393.37 nm) with the Balmer H$\delta$
and H$\zeta$ lines. He excluded the Ca~\textsc{ii} H line at 396.9 nm  
as it was blended
with H$\epsilon$ in the very low dispersion spectra he used (1.9 mm from
H$\gamma$ to H$\zeta$). With an exposure time of 90 minutes he could measure
spectra down to the 10th photographic magnitude, finding that for
the actual members of the Pleiades ``\textsl{there is a strict relation 
between the brightness and the spectral type, with no exception in the
interval from the 3rd to the 9th magnitude.}''. Hertzsprung's diagrams
(magnitude \textsl{vs} colour) of the Pleiades and the Hyades 
would appear a year later \citep{hertz1911} while
 Russell's version for field stars with
parallaxes  
 (with absolute magnitude \textsl{vs} spectral type)  
 would only appear in 1914 \citep{russell14a,russell14b}, although the 
correlation between
luminosity and spectral type was noted by  Hertzsprung in 1905 and in more detail by 
\cite{russell1912}. The key difference is that while Russell required parallaxes
to ascertain the distances to the stars, Rosenberg carefully checked the
membership of stars in the Pleiades, rejecting outliers and non-members.

Henry Russell was obsessed by priority, (self-)attribution and promotion 
\citep{devorkin2000}. For example, the famous
Vogt(--Russell) theorem on stellar structure first appeared in 1926 \citep{vogt1926}
and in his influential textbook Russell does give full credit to Vogt \citep{rds}, yet
he will  later claim \citep{russell1931} that he had found it independently. In the
case of the CMD, Russell called it in private the ``Russell diagram'', but this
was not accepted in public, as the contribution by Hertzsprung (unlike Rosenberg's)
was well and widely known. Russell poured over the astronomical journals, and was well aware
of Hertzsprung's results. We know he read the \textsl{Astronomischen Nachrichten} systematically,
as one of the leading journals of the time, and that we was well aware of Hertzsprung's
papers just as was his mentor, E.C. Pickering, who received them and wrote to  Hertzsprung
discussing several issues in spectral classification. Russell wrote to Hertzsprung
on September 27, 1910, thanking him for sending copies of his papers \citep{hearnshaw}.
\cite{hertz1911}'s paper contained the CMDs of the Hyades and the Pleiades, citing explicitely the
previous --and pioneering-- work by \cite{rosenberg1910}. With the raising influence of
European (mosly Dutch) astronomers in the USA, the issue of 
the proper acknowledgement became
very serious and created frictions and debate within the community. After two decades,
 the ``Russell diagram'' became known as the  Hertzsprung-Russell diagram, thanks in part
to the influential conference delivered in 1933 by B. Str\"omgren  
at the meeting of the Astronomische
Gesellschaft, but much to the irritation of many, including Russell himself who even
refused to acknowledge that Hertzsprung had found (and coined the terms) `giant' and `dwarf' 
stars \citep{smith1977}.
The (proper) renaming of the diagram was a long battle which lasted
till the late 1940s, when S. Chandraskhar, advising the \textsl{Astrophysical Journal} 
and tired of the controversy, decided that
the standard nomenclature would be the ``H--R Diagram'' \citep{devorkin2000}. 
Rosenberg's pioneering contribution has been unfairly forgotten from the 
history describing the elaboration of the first CMDs 
\citep[see, \eg,][]{waterfield1956,nielsen1969,devorkin2000}.

Hans Rosenberg was born in Berlin on May 18, 1879, and
studied first in Berlin, under W. Foerster, and then in Strasbourg, with E. Becker, 
obtaining his Dr. Phil.
with a thesis on the period changes that $\chi$ Cygni underwent from 1686 to 1901. Interested in both
instrumentation and astrophysics, 
he moved to G\"ottingen to work with Karl Schwarzschild where he was to 
produce the first large survey of stellar temperatures estimated with
objetive prism spectra. For years,
the Rosenberg temperature scale will set the standard and will be widely
used, as well as his review on photoelectric photometry in the \textsl{Handbuch der
Astrophysik} \citep{rosenberg1929}.  His \textsl{Habilitation} thesis from
T\"ubingen in 1910 was on ``\textsl{The relation between brightness and spectral
type in the Pleiades}'', whose results were published in the above-cited  key 
paper in \textsl{Astr. Nach.} \citep{rosenberg1910}. 
His instrumental expertise allowed him to get spectra of comets Daniel (1907 IV) and
Morehouse (1908 III).
 From 1910 on, he worked at T\"ubingen first as Privatdozent, where he founded its
observatory in \"Osterberg,  becoming its director in 1912 and  professor
at the university in 1916 while serving in the army during World War I. He made with P. Goetz
the first photometric map of the Moon, and developed the use of photoelectric cells as
astronomical detectors in 1913.
 Moving to Kiel in 1925 he worked on solar
eclipses, developed direct measures of the colours of stars and had heavy teaching duties.
In spite of his position as professor at Kiel university, on April 1, 1933, three uniformed members
of the feared SA Nazi paramilitary group  came to his appartment and forced him 
to resign \citep{duerbeck2006}, but he would
have lost his position all the same with the racial laws enacted in 1935. He was immediately 
invited in 1934 by the University of Chicago to work at Yerkes Observatory
where he stayed three years, working on measures of the limb darkening
and colour indices in eclipsing binaries \citep[\eg,][]{rosenberg1936}.
In 1938 he was appointed director of Istanbul Observatory, where he
reorganised the teaching of astronomy at the University of Istanbul and
set up new priorities for the observing campaigns. He passed away there
on July 26, 1940, a week after suffering a heat stroke \citep{turkref}.

An even earlier relationship between the colour and the magnitude was
found by \cite{charlier1889}, who studied the correlation between magnitude and a 
colour-like quantity (the difference between the  magnitude
he measured in a photographic plate and the visual magnitude) as
determined by Max Wolf in Heidelberg. However, he interpreted the correlation
found (the visually fainter stars had larger differences) as a systematic
error in the visual magnitudes. As Rosenberg correctly pointed out, this could also
be produced by selective reddening, hence the importance of getting spectra.
The spectral type -- magnitude correlation found in the Pleiades could not
be produced by dust, hence ``\textsl{the plausible colour differences
among the stars in the Pleiades --the fainter the star, the redder
it is-- following from the optical and photographic brightness measurements
 are confirmed by the spectral properties}''. This ground-breaking result
will be taken to good use to lay the foundations of stellar 
physics \citep[see, \eg,][]{salariscassisi2005},  but the true pioneer 
has unfairly been forgotten to the extent that it would be a fitting
tribute to rename the diagram as the Rosenberg-Hertzsprung-Russell
diagram (RHR). Ironically, HR also stands for Hans Rosenberg's initials.

The final word may come from Hertzsprung himself. His modesty made him to 
avoid talking about his own contributions to astronomy, and, as \cite{strand1968}
reminds us, he remarked on the controverted issue of the naming of the diagram:
\begin{center}
\begin{minipage}[t]{10cm}
\begin{quote}
``\textsl{Why not call it the colour--magnitude diagram? Then we know what it
is all about.}''\\
\end{quote}
\end{minipage}
\end{center}

The use of CMDs to constrain the physical properties of the stars was 
noticed very
quickly. In fact, \cite{russell1912} was the very first to use the correlation  between absolute
magnitude and spectral type for (dwarf) stars with measured parallaxes to infer a
distance to the Pleiades of 500 light-years\footnote{The spectra of the fainter members of
 the Pleiades came from E.C. Pickering and A. Cannon, and no mention is made of
\cite{rosenberg1910}'s work nor the comprehensive survey 
by \cite{hertz1911}.}.

\section{Single stars}
\label{dvg:single}
The fitting of isochrones to a set of stars is the main method to constrain their
physical properties, besides other techniques which are limited to nearby stars such
as stellar oscillations. The set of basic properties includes the age $t$, Helium
abundance $Y$, metallicity $Z$, $\alpha$ elements over abundance, distance modulus
$\mu = m - M$, convection mixing length, etc, that is, all the quantities which determine the position of
stars in a CMD. 
In contrast with its importance for stellar evolution, relatively little work 
has been done to formulate
mathematically the problem to go beyond the `fit-by-eye' approach which has
 characterised this field (and unfortunately still does!).
It seems that the first attempt to find the separation of a star from an isochrone comes
from  \citet{schaltenbrand1974} who developed a simple method to estimate the
nearest point of the zero-age main sequence to a given star in a two-colour diagram
under the assumption of Gaussian errors. This approach was further formalised in a
proper probabilistic framework by \citet{luri1992} as the so-called `proximity
parameter', similar to the deterministic `near point' estimator by \citet{flannery1982}. Here
one computes the sum of minimum distances from a set of $\Nstar$ stars to 
a given
isochrone with given properties (age $t$, Helium abundance $Y$, metal
content $Z$, etc) along with properties of the set of stars, which can
also be applied to the theoretical isochrone, such as the distance modulus
 $\mu$ or the extinction. We represent this set  as the
vector of parameters $\btheta=(t,Y,Z,\mu, \cdots)$. One can form the 
statistic $\Psi^2$ which is
calculated as the sum of minimal (squared) distances: 
\beq
\Psi^2(\btheta) \; = \; \sum_{i=1}^{\Nstar} \, \mathrm{min}_j
 \left( d_{ij}^2 \right) \quad ,
\eeq
where $d_{ij}$ is the geometrical distance between the observed star number $i$ and the 
theoretical point $j$
of the given isochrone:
\beq
d_{ij}^2 \; = \;  \left(\frac{m_i - m_j}{\sigma(m_i)}\right)^2 \; + 
\; \left(\frac{c_i - c_j}{\sigma(c_i)}\right)^2 \quad ,
\eeq
and $m$ and $c$ are magnitudes and colours respectively, or gravities and
temperatures, or any two observables which can be predicted with the models,
while $\sigma(m_i)$ and
$\sigma(c_i)$ are the errors in these quantities for the given star labeled $i$. 
This distance can obviously be generalised to any $n$-dimensional space (see below). 
If (and this is a major assumption, see below) the model stars were uniformly distributed along the isochrone, then the
probability that one observed star comes from that isochrone would be
\beq
\mathcal{P}_i(\btheta) \; = \; \frac{1}{2  \, \pi \, 
\sigma(m_i)  \, \sigma(c_i) } \; \sum_{j=1}^{N_{iso}} \, 
\exp\left( - d_{ij}^2 / 2 \right) \; ,
\label{dvg:eq1}
\eeq
where $N_{iso}$ is the total number of points the isochrone has been divided into. 
The probability that an \textsl{ensemble} of $\Nstar$ stars comes from an isochrone
with the set of properties $\btheta$ is just the product of the individual probabilities
\beq
\mathcal{P}_{total}(\btheta) \; = \; \prod_{i=1}^{\Nstar} \, 
\mathcal{P}_i(\btheta) \; .
\eeq
One can then use the standard maximum likelihood technique to find the set
of parameters $\btheta$ which maximises this probability. The maximum likelihood
statistic (MLS) would for instance be
\beq
\mathrm{MLS}(\btheta) \; = \; - \log \mathcal{P}_{total}(\btheta) 
\; = \; \sum_{i=1}^{\Nstar} \, \log \mathcal{P}_i(\btheta) \; .
\eeq
For, say, $m=3$ parameters (distance modulus, age, and metallicity) this would be an optimisation
problem: find the point in this 3-dimensional space for which $\mathrm{MLS}$ reaches
a maximum. 

\begin{figure}[!hpt]
\includegraphics[width=\textwidth]{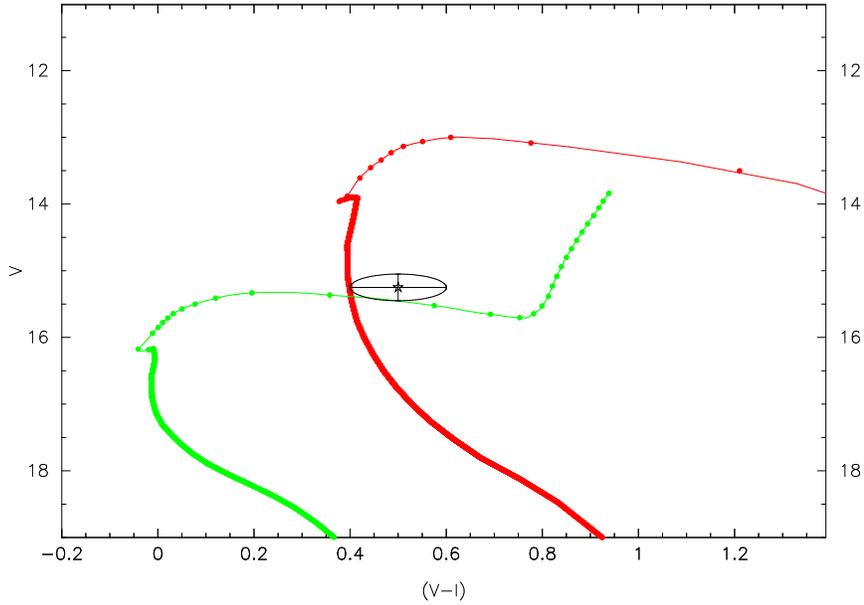}
\caption{The observation of a single star in a given CMD. Which of the
two isochrones is more likely to be the correct one? Only two isochrones 
are plotted, for illustrative purposes, with different properties (age, distance, metallicity, etc), and the
points along each isochrone are such that the diffence in stellar mass is
the same (namely $\Delta m = 0.01 \Msun$). In their main sequence the density of stars is very
large (slow evolution) while after the turn-off, the much faster evolutionary
speed makes the stars to become more widely separated for this
fixed $\Delta m$. Many different isochrones can
go through the same observed point (star with error bars at 1$\sigma$ and with the 
corresponding
probability level given by the elliptical curve), and hence there are, in principle, many possible
solutions: there is a huge degeneracy. For clarity, in this Figure only two 
isochrones are marked whose Near Point lie
at a normalised distance of 1$\sigma$ from the observed position. The geometrical term of their likelihoods is
identical (same distance in probability or $\sigma$ units), yet the lower (green) isochrone has an
evolutionary term much smaller than the
one from the upper (red) isochrone: the density of points from the lower (green) isochrone
is much smaller when arriving close to the observed star. 
Hence,  it is much less probable that the observed star is drawn from it.
The purely geometrical degeneracy can be lifted using the prior information
provided by stellar evolution in terms of the evolutionary speed of each track.
In addition, another prior comes from the observed stellar mass function, which also favours smaller 
stellar masses.}
\label{fig:dvg:single}
\end{figure}

The procedure is best illustrated for the case of a single star ($\Nstar=1$), as shown
in Fig.~\ref{fig:dvg:single} where a star (and its errors, which determine the
elliptical confidence region) is represented in a CMD along with two possible isochrones
of very different properties but which lie at the same normalised distance $d_{1m}$ 
of unity. Clearly many more isochrones can have the same normalised distance or
even smaller ones, but they both show that one cannot decide which isochrones fits
best, since both contribute the same amount to the $\Psi^2$ statistic or to
the   $\mathrm{MLS}$ one. The problem,
thus formulated, is intrinsically degenerate, as there are multiple solutions: many
different combinations of the basic parameters $\btheta$ can give the same
maximum in the likelihood.

The reason for this degeneracy stems from the key assumption made in deriving the
probability that the star was sampled from the isochrone (Eq.~\ref{dvg:eq1}): it was assumed
that the number of stars \textsl{along} the isochrone was the same. That is, we
were only concerned about the \textsl{geometrical shape} of the isochrone, and not about
the \textsl{density} of stars along it. This \textsl{geometrical method}, while useful
in selecting shapes of isochrones which come close to the observed position, is
highly degenerate (Fig.~\ref{fig:dvg:single}).

\begin{figure}[!ht]
\includegraphics[width=\columnwidth]{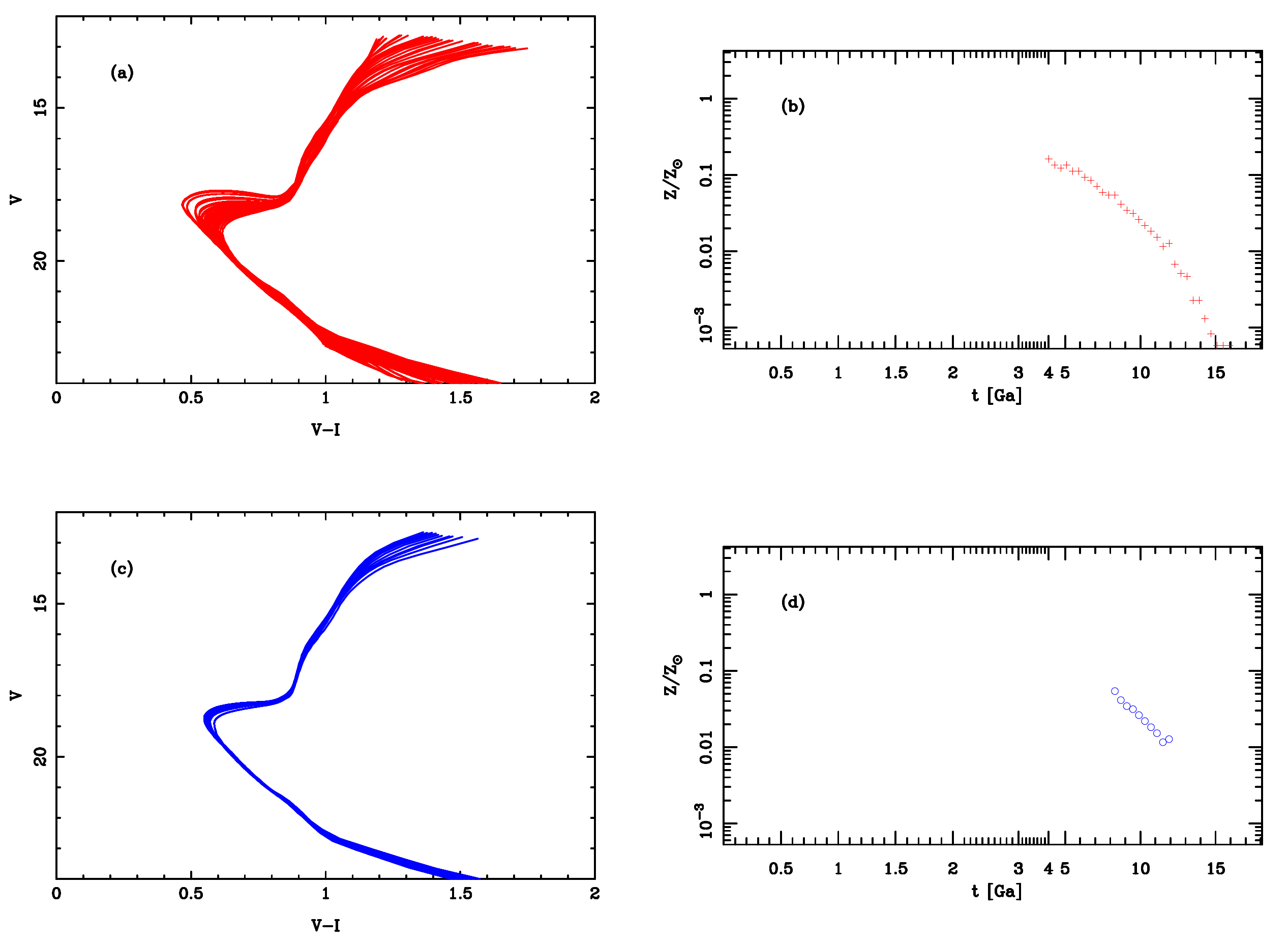}
\caption{An illustration of the \textsl{geometrical} degeneracy between age and
metallicity. In panel (\textsl{a}) a series of isochrones of roughly similar
shapes is indicated, whose ages and metallicities are given in panel
(\textsl{b}). The well-defined locus in the $Z$-age plane means that many
CMD inversions with sparse populations, differential reddening or binaries
will tend to produce age-metallicity ``relations'' which only reflect this
geometrical degeneracy. This is further illustrated in panel (\textsl{c}) where
the geometrical shapes are even closer. Yet, as panel (\textsl{d}) shows, any
isochrone changing its metallicity by nearly an order of magnitude can produce
another isochrone of the very same geometrical shape provided the age changes
by about a factor of two. Physically, there is no such degeneracy, as a change
of (atmospheric) metallicity reflects a change in the (core) abundance, and
hence a different evolutionary speed, as reflected by the density of stars
along the isochrone.
}
\label{fig:dvg:degenerate}
\end{figure}

Figure~\ref{fig:dvg:degenerate} illustrates one aspect of this degeneracy: isochrones
of widely different ages and metallicities have a very similar geometrical
shape. It is therefore hardly surprising that geometrical methods which
rely on the proximity of a star to these isochrones yield huge degeneracies.
Adding the uncertainties in distance modulus and dust extinction makes these
methods unsuitable for any quantitative analysis.

While extensions of this
geometrical method are discussed in their contexts later on, it is worth
understanding in more detail the underlying reason for which the distribution
of stars is \textsl{not} uniform along an isochrone.

Let us consider a curvilinear coordinate $s$ along an
isochrone of given parameters. Stellar evolution theory predicts that for
a given abundance the
number of stars on the isochrone  
depends only on the age $t$ of that isochrone
and on the mass $m$ of the star at that precise locus, so that $s=s(m,t)$ only. 
An offset in
age $dt$ is reflected only through changes in mass $m$ and position $s$ along 
the isochrone of
age $t$ since
\beq
dt(m,s) \; = \; \left.\frac{ \partial t }{\partial m}\right|_s \, dm \; + \;
             \left.\frac{ \partial t }{\partial s}\right|_m \, ds \qquad .
\eeq
For a star in the isochrone the offset is, by definition, $dt = 0$ hence
\beq
\left.\frac{ \partial m }{\partial s}\right|_t \; = \; - \;
\left.\frac{ \partial m }{\partial t}\right|_s \; \times \;
\left( \left.\frac{ \partial s }{\partial t}\right|_m \right)^{-1}   \quad .
\label{dvg:eq2}
\eeq
The first term is always finite.
The second term is the evolutionary speed, the rate of change for a given mass $m$
of its coordinate along the isochrone when the age changes by some small amount. One can
think of $s$ as representing some evolutionary phase, and so this term will be large
when the phase is short-lived : a small variation in age yields a very large
change in position along the isochrone. Alternatively, for a given age, and since
the first term is always finite, a wide variation in position implies a narrow
range in mass. This is the case of the red giant branch or the white dwarf
cooling sequence, for instance. On the other hand, slowly evolving phases such
as the main sequence have small evolutionary speeds and wide ranges in mass
for a given interval along an isochrone. Clearly, the most important phases to
discriminate between alternative ages and metallicities will be the post main-sequence
ones, where the range of mass is small (and hence insensitive to the details of the
stellar mass function), and at the same time where evolutionary speeds are large.
If we consider the mid- to lower main sequence, at fixed metallicity, isochrones of all ages trace the same locus,
with only marginal changes in the density distribution of points amongst them. In this sense, one
of the parameters, the age $t$, is to a large extent absent from the main sequence,
while phases beyond it are always substantially a function of (at least) 
both age and metallicity.

The density of stars along an isochrone is therefore
\beq
\frac{d N}{d s} \; = \; - \left(\frac{dN}{dm}\right) \times
\left( \left. \frac{\partial m}{\partial t}\right|_s \right) \times
\left( \left. \frac{\partial t}{\partial s}\right|_m \right) \qquad .
\label{dvg:eq:density}
\eeq
The first term is related to  the initial stellar mass function (IMF), and, as discussed above,
the second term is finite while the third is a strong function of the evolutionary
speed. If the mass after the turn-off is assumed to be roughly constant, this
implies that the {\sl ratio} in the number of stars in two different evolutionary
stages after the turn-off will only depend on the ratio of their evolutionary
time scales. In the context of stellar population synthesis, this is
known as the fuel consumption theorem \citep{renzini83}.

The way to infer the properties of star, given some observables, and our knowledge of
stellar evolution, is obviously the Bayesian method, where both the errors in the
observables and our prior information (stellar evolution) can be handled properly, 
even in the case of one single star. Good
reviews of Bayesian inference in physics are provided by \citet{cousin1995}, 
\citet{dose2003} and \citet{trotta2008}, 
 along with the monographs by \citet{gregory2005} and \citet{hobson2010}. The Bayesian method allows us to answer
the question we are interested in: \textsl{ What is the probability 
of the occurence of the estimated parameters $\btheta$ (mass, distance, age, etc) 
given the observed dataset $\bdata=(V, V-I, B-R,  \cdots)$ and our prior information provided by stellar evolution?} The
prior information can be formulated as the probability distribution function expected for the
parameters, $\pi(\btheta)$, on the basis of stellar evolution or
other prior knowledge (say, a measure of $Z$), and is normalised to one, 
$ \int \pi(\btheta) \, d\btheta = 1$. The probability of observing the dataset $\bdata$, given some
parameters $\btheta$ is the likelihood $\like(\bdata|\btheta)$. An important quantity is the 
\textsl{evidence} (sometimes also termed as marginal likelihood) which is just 
$E(\bdata) \; = \; \int \pi(\btheta) \,  \like(\bdata|\btheta) \, d\btheta$. \citet{bayes1763}'s 
theorem\footnote{\citet{dale1982} explores the issue of whether \citet{laplace1812} should rather
been given credit to the actual use, proof and development of the theorem.} 
then states that 
\beq
{\mathcal P}(\btheta|\bdata) \; = \; \pi(\btheta) \; \frac{ \like(\bdata|\btheta) }{E(\bdata)} \qquad .
\eeq
In other terms, our prior information of the parameters is modified into our posterior probability
distribution function by the ratio of the likelihood over the evidence. We will see that this
formulation of the problem also allows us to discriminate among models through the model
selection technique. In the case of an observed star to be associated with an isochrone with
$n$ observables $\bdata$ and $n'$ parameters $\btheta$, we can define a simple \textsl{geometrical} likelihood as
\beq
\like_{geom}(\bdata|\btheta) \; = \; \frac{1}{ (2 \, \pi)^{n/2} \, \prod\limits_{k=1}^n \sigma_k} 
\; \prod\limits_{k=1}^n \, 
\exp\left( -D_k^2 /2 \right) \qquad ,
\label{dvg:eq:likegeo}
\eeq
where $D_k$ is again the normalised distance between the observed quantity $S_k$ and the one
predicted by the model $M_k$, given an observed error $\sigma_k$ in that quantity 
\beq
D_k^2 \; = \; \left( \frac{ S_k - M_k }{\sigma_k} \right)^2 \qquad .
\label{dvg:eq:distance}
\eeq
For example, we can take $n=3$ observables such as 
$\bdata=(m_V^{obs}, T_{\mathrm{eff}}^{obs}, Z^{obs})$, 
which depend on, say, $n'=5$ theoretical parameters such as $\btheta=(m, t, d, Z, \alpha)$ and we
would have
\begin{eqnarray}
\like_{geom}(m_V^{obs}, T_{\mathrm{eff}}^{obs}, Z^{obs} | m, t, d, Z, \alpha) \;
 & \;  =  \nonumber \\
  \frac{ \exp\left( - \chi^2 / 2 \right)}{ 
(2 \, \pi)^{3/2} \; \sigma(m_V^{obs}) \; \sigma(T_{\mathrm{eff}}^{obs}) 
 \; \sigma(Z^{obs})} \;  & \; ,
\end{eqnarray}
with, under the assumption that they are uncorrelated, 
\begin{eqnarray}
\chi^2 & \; = \; & \left( \frac{ m_V^{obs} - m_V^{theo}(m, t, d, Z, 
\alpha)}{\sigma(m_V^{obs})} \right)^2 \nonumber \\
       & \; + \; & \left( \frac{ T_{\mathrm{eff}}^{obs} - T_{\mathrm{eff}}^{theo}(m, t, d, Z, \alpha)}
{\sigma(T_{\mathrm{eff}}^{obs})} \right)^2 \nonumber \\
       & \; + \; & \left( \frac{ Z^{obs} - Z^{theo}(m, t, d, Z, \alpha)}{\sigma(Z^{obs})} \right)^2 \quad .
\end{eqnarray}
If  quantities are correlated, one has to form the pairs
\beq
\like_{corr}(A,B) \; = \;  \frac{ \exp\left( - 
\left[ D_A^2 + D_B^2 - 2 \rho D_A D_B \right] / \left[ 2 (1-\rho^2)\right] \right)}{ 
2 \, \pi \; \sigma(A) \; \sigma(B)  \; \sqrt{1 - \rho^2} } \;  , 
\eeq
where $\rho$ is the correlation coefficient between the two quantities $A$ and $B$, or,
in terms of their covariance, $cov(A,B) \, = \, \rho \, \sigma(A) \, \sigma(B)$.

We can illustrate this case with an example where we assume to have a prior knowledge of 
the distance of the star
and its bolometric correction so that we can for instance use the 
bolometric luminosity rather than
its apparent magnitude in some photometric band.

\begin{figure}[!ht]
\includegraphics[width=\textwidth]{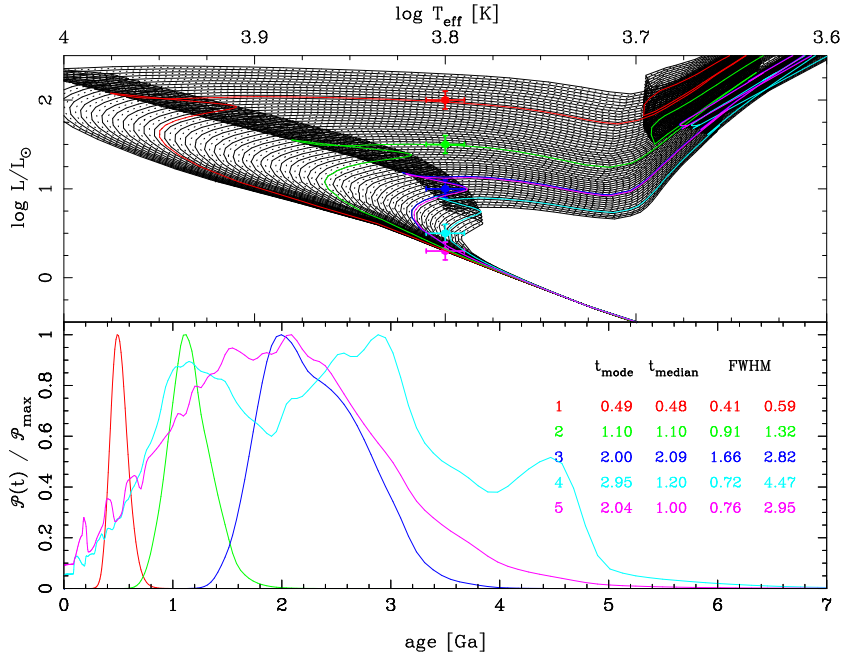}
\caption{A test case with 5 stars at fixed effective temperature ($\log T_{\mathrm{eff}}
=3.8$, $\sigma(\log T_{\mathrm{eff}}) = 0.01$) and five different bolometric
luminosities
sampling the HRD in regions where: (\textsl{i-ii}) evolution is fast (top two stars, colour-coded
red and green); (\textsl{iii}) near the TAMS, where multiple isochrones cross (blue
star ($\log L/L_{\odot} = 1.0$);
(\textsl{iv}) a slightly evolved star away from the MS (pale blue); and
(\textsl{v}) a star very close to the ZAMS traced by very young stars. The
lower planel shows the posterior probability distribution function for the age
of each star, assuming flat priors for the other parameters.
 In this figure, the likelihood is computed using the geometrical
term \textsl{only}, and while the upper two stars appear to have well-defined
ages (a FWHM range from 0.41 to 0.59 Ga for the first star, and from
0.91 to 1.32 Ga for the second one), the star located between the red and the
blue loops has a FWHM range of 1.5 Ga, and it gets worse for the two lower
stars. This is just due to the large density of isochrones in these areas,
and the geometrical term of the likelihood cannot disentangle, per se, which
set is more appropriate, since no prior information on the evolutionary
speed is used.
}
\label{fig:dvg:hrdgeo}
\end{figure}

Fig.~\ref{fig:dvg:hrdgeo} shows as a test case five stars with the same effective temperature, but
different luminosities so as to sample different evolutionary r\'egimes (top panel). There is
no prior information on age or metallicity nor mass. If we are only interested in, say, the
ages of these stars, we can consider the mass, the metallicity, the
mixing length parameter, etc, as nuisance parameters with
flat probability distributions (say, between 0.2 and 200 $\Msun$, 0.0001 and 0.2, 
and 0.5 to 2.0, respectively)
 so that the probability distribution function (PDF) of the age is given by
\beq
\mathcal{P}(t) \; = \; \int \, \int \, \int \, dm \, dZ \, d\alpha \; \like_{geom}(t, Z, m, \alpha) \quad .
\eeq
Likewise, if we are interested in estimating the mass of that star, we would marginalise
over the other (nuisance) parameters to get
\beq
\mathcal{P}(m) \; = \; \int \, \int \, \int \, dt \, dZ \, d\alpha \; \like_{geom}(t, Z, m, \alpha) \quad .
\eeq
We illustrate the technique here for the ages of the stars, given their importance in the context
of both CMDs and stellar evolution \citep[\eg,][]{lebreton2000,soderblom2010}, and because
they have been widely used, even though they are not physically justified 
\citep[e.g][]{lachaume1999,reddy2003}.
The resulting age PDFs are given in the lower panel of Fig.~\ref{fig:dvg:hrdgeo} and show the
widely different distributions depending on the location of the test stars in the diagram (all
share the same errors in bolometric luminosity and effective temperature, for the sake of
the argument).

More importantly, however, this formulation of a \textsl{geometrical} likelihood (Eq.~\ref{dvg:eq:likegeo})
does \textsl{not} include any information on the evolutionary speed: we have only used the geometrical
\textsl{shape} of the isochrone, the only quantity that
matters when computing the distance from the observed star to a point on
the isochrone (Eq.~\ref{dvg:eq:distance}). To incorporate the \textsl{physics} of stellar evolution we need to account
for the \textsl{density} of stars along the isochrone (Eq.~\ref{dvg:eq:density}) so that we have
a proper \textsl{physical} likelihood. To do this, we need to integrate along all possible
masses in the isochrone, but noting that (unlike the geometrical case) not all masses
are equally probable. Let the density of stars of mass $m$ along an isochrone be $\rho(m)$,
then 
\beq
\like_{phys}(\bdata|\btheta) \; = \; K^{-1} \; \int\limits_{m_{lim}}^{m_{top}} \; dm \; 
\frac{ \rho(m) }{ (2 \pi)^{n/2} \prod\limits_k^n \sigma_k} 
\; \prod\limits_k^n \, 
\exp\left( -D_k^2 /2 \right) \quad ,
\label{dvg:eq:likephys1}
\eeq
where, for a proper normalisation, we require
\beq
K \; = \; \int\limits_{m_{lim}}^{m_{top}} \; dm \; \rho(m) \quad , 
\eeq
and $m_{lim}$ and $m_{top}$ are the lower and upper mass limits of the isochrone
considered. 
It is useful to write this in terms of the curvilinear coordinate $s$ along the
isochrone, since the mapping of the mass $m$ to a position in the CMD is highly
non-linear (\cf Eq.~\ref{dvg:eq:density}): the red giant branch will be poorly sampled if the mass interval is too
large. We can thus re-write Eq.~\ref{dvg:eq:likephys1} as
\beq
\like_{phys}(\bdata|\btheta) \; = \;  \; \int\limits_{s=0}^{s=1} \; 
\underbrace{ds}_{curvilinear} \; 
\underbrace{\frac{dN(m)}{dm}}_{\phi(m)} \; \; \underbrace{\frac{dm}{ds}}_{speed} \; \;  
\underbrace{\frac{\prod\limits_k^n \, 
\exp\left( -D_k^2 /2 \right)  }{ (2 \pi)^{n/2} \prod\limits_k^n \sigma_k}}_{geometry}
\;  \quad ,
\label{dvg:eq:likephys2}
\eeq
where the lower and upper limits of the curvilinear coordinate $s$ have been set to
0 and 1 respectively, and we can identify, besides the geometrical term, the initial
mass function $\phi(m)$ and the evolutionary speed $dm/ds$ along the isochrone. These
two functions encapsulate the prior information provided by stellar evolution, in
a way that the geometrical approximation cannot possibly handle. In reference to
previous works \citep{Hernandez:1999lo,bjarne}, we will refer this method as the
\texttt{BayesGM} method\footnote{In \citet{Hernandez:1999lo} the physical likelihood
was referred to as the G matrix.}. 

The posterior PDF then becomes
\beq
\mathcal{P}_{post}(\btheta|\bdata) \; = \; \prod\limits_{k=1}^{n} \; \pi(\btheta)_k 
\; \prod_{i=1}^{\Nstar} \, \like_{phys}(\bdata|\btheta)_i \; ,
\label{eq:dvg:post}
\eeq
where $\pi(\btheta)_k$ is the prior probability distribution function for the
parameter indexed $k$ of the parameter vector $\btheta$.  The key point of this 
formulation is that the prior on mass, the IMF $\phi(m)$, must be \textsl{included} in
the likelihood, for a proper physical weighting. 
As the dataset is
fixed, we do not need to compute the evidence which only acts, in this
context, as a normalisation
constant. 

\begin{figure}[!ht]
\includegraphics[width=\textwidth]{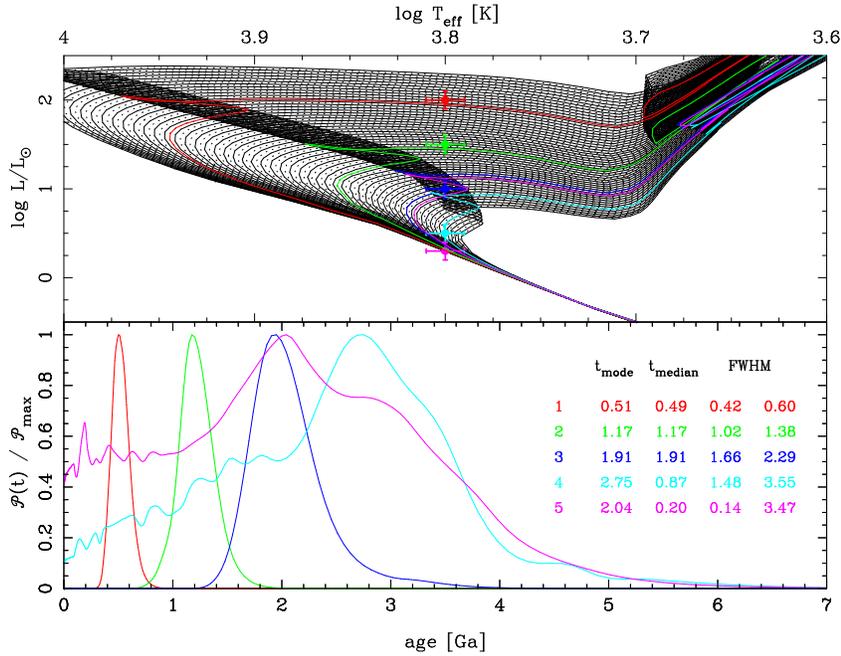}
\caption{The same set of stars as in Fig.~\ref{fig:dvg:hrdgeo}. This time
the full likelihood function (that is, with the evolutionary terms) and the
prior on the mass function is used. There is a tendency to increase the mode
and the median age for the two top stars, although still within the same
FHWM range. This slight increase is due to the fact that the probability of
observing a less massive star is larger, and hence the ages increase. This
effect helps reducing the range for star \# 3 by almost 0.5 Ga, and produces
well-defined peaks for the two lower stars, reducing the range of
 the posterior PDF. The effect is even stronger when a wide range of
metallicities is considered: the posterior PDF peaks at the proper ages and
metallicities, hereby lifting the purely geometrical degeneracy illustrated
in Fig.~\ref{fig:dvg:degenerate}.
}
\label{fig:dvg:hrdphys}
\end{figure}

For example, consider a star on the red giant
branch: for a long interval along the nearly vertical isochrone, the mass hardly
changes, and hence the effective weight of the IMF in the integrand will be very
small and the geometrical distance will, in proportion, be more important. In the main sequence the
reverse is true. The effects of including explicitely in the likelihood 
this prior information are dramatic, and shown in Fig.~\ref{fig:dvg:hrdphys}. There is
little difference for the two brightest stars, as isochrones run almost parallel to
the effective temperature axis, and hence map the plane in a well-behaved way: the
only difference with the geometrical method is that our prior information on both
evolutionary speed and the IMF will favour tracks with \textsl{smaller} masses (they
evolve more slowly and are more abundant, two factors that cannot be handled by the
geometrical method). The middle star (colour-coded in blue) is more interesting as
it lies in the area where multiple isochrones cross each other, hence providing
a huge geometrical degeneracy as reflected by the age range from 1.66 to 2.82 Ga 
 (Fig.~\ref{fig:dvg:hrdgeo}). In
this particular case, large ages appear to be penalised, and the new FWHM range is
restricted to 1.66 to 2.29 Ga (Fig.~\ref{fig:dvg:hrdphys}). The next star illustrates this effect even more clearly,
with a reduction of the interval 0.72--4.47 Ga to 1.48--3.55 Ga, a contraction of
1.75 Ga. The faintest star, which appeared deceptively well behaved when using the
geometrical method, now reveals that much younger ages are twice more probable, as
well as older ones. The full prior information on the way stars evolve is properly
used.

In Figs.~\ref{fig:dvg:hrdgeo} and ~\ref{fig:dvg:hrdphys} the full posterior
age distributions are given. Rather than using the full PDF of the parameters, 
it is customary to encapsulate the
information in quantities such as the (posterior) modes (the values of the parameters
 where the posterior PDF reaches a maximum), the means or expectation values, or even
the medians. The frequentist confidence intervals become, within the Bayesian framework,
the credible regions (CRs) such that they are the (closed but not necessarily connected) 
volumes which contain a fraction $\alpha$ of the total volume under the posterior: 
\beq
\int_{CR(\alpha)} \, d\btheta \; \mathcal{P}_{post}(\btheta|\bdata) \; = \; \alpha \; \; .
\eeq
There are many different possible CRs. The central credible interval (CCI) is defined
such that the intervals $(-\infty,\theta_{low})$ and  $(\theta_{high},+\infty)$ each
contains $(1-\alpha)/2$ of the posterior volume, and always contains the median. The
minimum credible interval (MCI) is built in such as way tha the posterior PDF is always
larger inside the MCI than outside. It contains the mode, obviously,  but may not
be connected. In Fig.~\ref{fig:dvg:hrdphys} the FWHM (that is, the values for which
the PDF reaches half its maximum value) are indicated, but note that they do not
correspond to any fixed probability measure. The figure also illustrates the poor
performance of the median when the PDF is wide. Similarly, and contrary to the claims
by \cite{burnettbinney2010}, the mean can be highly biased (although easy to compute). We
hence prefer to use the mode of the distribution, as a robust point estimate of the
quantity of interest. 
 
\begin{figure}[!ht]
\includegraphics[width=\textwidth]{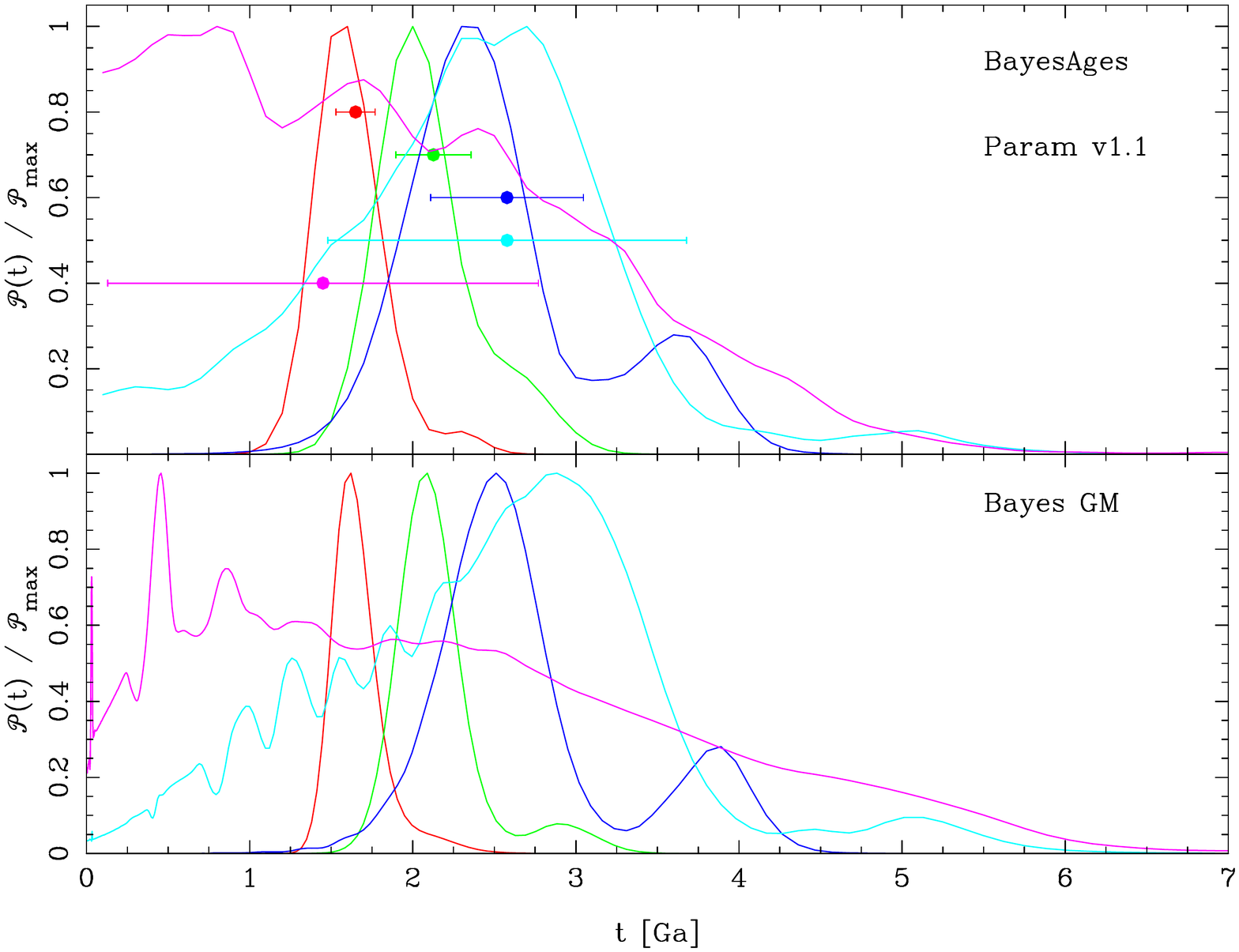}
\caption{Comparing the posterior PDF in ages for the 5 stars indicated
in the top panels of Figs.~\ref{fig:dvg:hrdgeo} and \ref{fig:dvg:hrdphys},
with uncertainties of $\sigma(\log T_{\mathrm{eff}}) = 0.01$ and $\sigma(M_V) = 0.1$,
this time at $M_V = 2.0, 2.5, 3.0, 3.5$ and 4.0.
\texttt{BayesAges} \citep{ponteyer2004} tends to produce slightly wider
PDFs than our \texttt{BayesGM} technique, while \texttt{PARAM v1.1} \citep{dasilva2006} 
yields point estimates slightly older (1$\sigma$ bars have been offset vertically for
clarity). The well-defined peak at $t \approx
0.45$ Ga predicted by \texttt{BayesGM}, and which is not present in
\texttt{BayesAges} is robust: decreasing the uncertainties in the two
observables produces a mode at this age.
}
\label{fig:dvg:comparison}
\end{figure}

\cite{ponteyer2004} have also developed a Bayesian technique to explore the distribution
of ages for the particular case  of G dwarf stars to revise the age-metallicity 
relation in the Galactic disc. They use a \textsl{ 
geometrical} likelihood in the 3-dimensional parameter space of effective temperature,
absolute magnitude and metallicity, and include stellar evolution is terms of
priors (for instance, taking the variation of the luminosity as a function of age
at fixed temperature). Fig.~\ref{fig:dvg:comparison} compares the results of their
method, \texttt{BayesAge}, with \texttt{BayesGM} for five test stars. While, as
expected, the results are very similar for the brightest stars, for the fainter ones
the bias they observe towards younger ages is not present in \texttt{BayesGM}. For
the faintest test star, the peak predicted by \texttt{BayesGM} is confirmed when
decreasing the error bars in the observables, confirming the robustness of the method.
\cite{dasilva2006} use a physical likelihood but take the IMF as describing the number of stars along
a given isochrone, which is not correct (except on the ZAMS, see Eq.~\ref{dvg:eq:density}). Nevertheless, the estimates obtained
agree with \texttt{BayesGM} while they seem to be systematically larger than \texttt{BayesAges}, as
indicated in Fig.~\ref{fig:dvg:comparison}. The code \texttt{PARAM v1.1} is available at 
\url{stev.oapd.inaf.it/cgi-bin/param}. 
Remarkably, \cite{valenti2005} did use what we could call a `empirical' bayesian technique
to weigh geometrically-estimated ages through a variety of `probabilities' in a purely
empirical way. In contrast, 
\cite{takeda2007} used a proper Bayesian formalism which includes the possible variation of $\Delta Y / \Delta Z$
in stellar tracks, and limit the integration of the posterior PDF to the hyper-box in the
space of parameters.

\cite{breddels2010} apply a maximum likelihood technique again using a purely
geometrical criterion, while \cite{burnettbinney2010} develop a Bayesian method
to infer the first two moments of the posterior PDFs, again with a purely geometrical
likelihood but properly weighted by priors. In contrast, \cite{casagrande2011}, building upon their previous
work \citep{casagrande2010}, include explicitely priors which attempt to correct
for biases known to exist in their sample and a proper physical likelihood.
\cite{bailerjones2011} also uses a Bayesian framework to estimate extinction and stellar parameters
from measures of parallaxes and multi-band photometry. The prior information from stellar evolution
is included explicitely by a distribution obtained with 200,000 stars of solar metallicity 
sampled from a Salpeter IMF
and a flat  star
formation history. The (smoothed) density (obtained via a Gaussian kernel) is used as a prior in absolute
magnitude and temperature. 

Isochrone ages have now the statistical framework to be inferred properly, and which
can be used to calibrate gyrochronological ages \citep[\eg,][]{chaname2012}, and be
compared with the independent measures obtained through 
asteroseismology \citep[\eg,][]{stello2009}. The field has evolved 
dramatically since the first attempts \citep[\eg,][]{perrin1977} with purely empirical
fits to massive surveys with proper statistical
methods \citep[\eg,][]{nordstrom2004}. The advent of large-scale surveys with
measures of fundamental parameters, such as \textsl{RAVE} \citep[\eg,][]{zwitter2010} and \textsl{GAIA} 
will show the usefulness of these Bayesian techniques.

\subsection{Caveats and (some) systematics}

There are two important caveats worth keeping in mind, besides the ones
noted in \S~\ref{dvg:sec1} (photometric corrections, 
bolometric corrections and standard values) as there are two
underlying assumptions that have been made: first that the light received
in the detector actually corresponds to the star, and second that the position
of the star in the CMDs  is only determined by its mass, for 
a given age and metallicity. The first assumption could be wrong if the
detected star is, in fact, an unresolved binary or multiple system. In this case, we are
detecting the combined light from the components of the 
system, and both the magnitude and the colour are shifted by an amount
which depends on the mass ratio(s).  In the
theoretical diagram, the luminosity and effective temperature of the combined system
is related to the individual properties as
\begin{eqnarray}
L_{A+B+C+\cdots} &\; = \; &L_A + L_B + L_C + \cdots \quad ,\\
T_{\mathrm{eff}(A+B+C+\cdots)}^4 & \;  = \; & \frac{L_A + L_B + L_C + \cdots}{4 \pi \sigma \, 
(R_A^2 + R_B^2 + R_C^2 + \cdots )} \quad , 
\end{eqnarray}
and not, as some authors have wrongly claimed \citep{siess1997}, as the luminosity-weighted
mean of the effective temperatures. 
In the CMDs this effect gives rise to well known offsets from single star sequences 
\citep[\eg,][]{haffner1937,maeder1974,lastennet1996,hurley1998} 
 and unless one has good reasons to assume the star in consideration is truly single, the
effect will produce a systematic bias in the inferred properties.

The second assumption covers several effects. 
The first one is just a consequence of \citet{vogt1926}'s theorem, which is
not valid for instance in cooling white dwarfs, where the complicated
details of the cooling sequence no longer depend only on the mass. In
principle, the parameters controling the locus of the white dwarf in
the cooling sequence could be included in the multivariate set $\btheta$,
and the same formalism can be applied \citep{vonhippel2005,vonhippel2006,
jeffery2007,vandyk2009,jeffery2011}. The same applies to pre-main-sequence stars,
whose tracks depend to a large extent on their mass accretion history, which
introduces a major complication, with, in principle, a functional degree of
freedom which is difficult to constrain \citep{maynenaylor,naylor2009}. Attempts 
are currently being
made to use a Bayesian formalism to tackle this problem \citep{gennaro2012}.

An equally serious case where the assumption is known to be wrong is
provided by massive stars, whose fast rotations not only bring fusion
products from the core to their surface during core hydrogen burning,
and hence affect the abundances, but also their location in a CMD depends
at least both on the inclination and on the equatorial velocity, none
of which are measurable (only $v \sin i$ can be inferred from the
line profiles). The effects of rotation can reach some 0.1 mag or more
in both colour and magnitude \citep{maeder1970,collins1977}. 
While statistical techniques \citep[\eg,][]{collins1985,lastennet1996} could be used 
incorporating further parameters into the $\btheta$ vector, the
availability of 
state-of-the-art models of (massive) rotating stars  \citep[\eg,][]{brott2011,maeder2012}
may provide another way of dealing with stars in the upper main sequence or beyond.

The third effet is the assumed enrichment $\Delta Y / \Delta Z$ which is built-in in 
the evolutionary tracks. Clearly different assumptions on both the Helium abundance
 $Y$ and the enrichment ratio have consequences on the stellar speed, and thus far only
tests with two different sets have been carried out \citep{casagrande2011}, but much
more tests need to be done.

Last, but not least, is the thorny issue of the calibration of the mixing length  
convection theory (MLT): all tracks/isochrones  are normalised to the putative solar case,
and so stars with widely different masses and metallicities are still assumed to
have the very same properties as the solar convective layers. While there is some mild evidence for a possible
variation of the MLT parameter (which describes the mixing length in units
of the local pressure scale) with mass in some binary systems 
\citep[\eg,][]{lastennet2003,yildiz2006,yildiz2007} no isochrones have so far
been computed for different scalings of the MLT parameter with mass and/or metallicity,
and yet it is clear that stellar convection must depend on stellar parameters, as
3-dimensional simulations are indicating \citep{ludwig1999}. Similarly the
amount of overshooting is yet another degree of freedom which is rarely
taken into account even if tracks are computed for a variety of possible
values \citep[\eg,][]{vandenberg2006}.

All these provisos are worth keeping in mind before any interpretation of the
resulting PDFs  is carried out.

\section{Detached binary stars}
\label{dvg:double}
Binary stars can provide, in some circumstances, the only direct and
reliable way to measure stellar masses and hence  constitute a benchmark
for stellar evolution. They can also provide accurate measures of many
other quantities, and in some cases the full orbital and physical
parameters. Astrometric (or interferometric) orbits combined with
radial velocities, or detached eclipsing binaries which are also
double-lined spectroscopic ones yield a wealth of precise measures  
reaching sometimes 1\% in masses and radii 
\citep[see][for a comprehensive review]{tag2010}. 
A catalogue of over 130 well-measured  binary systems is 
mantained at 
\url{www.astro.keele.ac.uk/jkt/debcat}.

\begin{figure}[!hb]
\includegraphics[width=\textwidth]{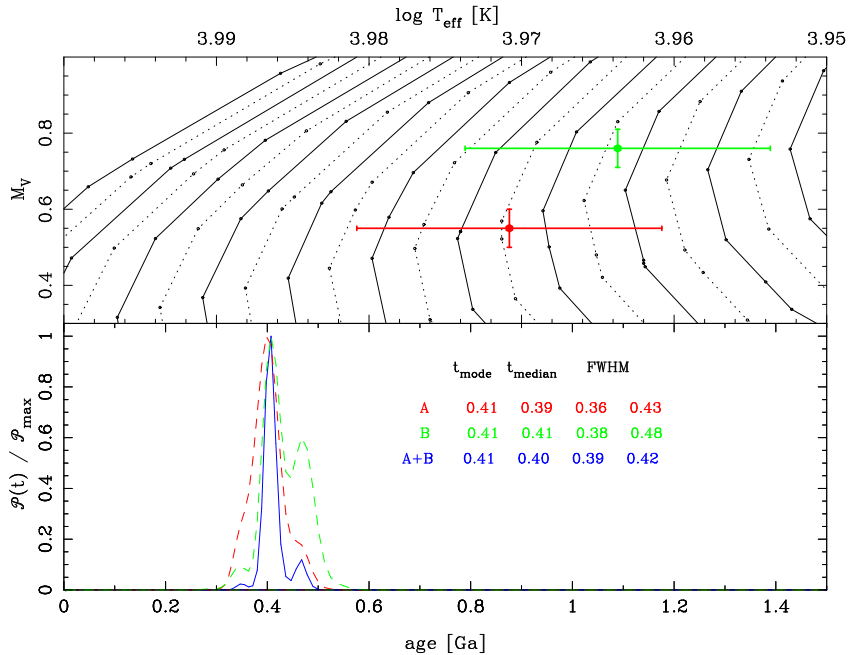}
\caption{Testing Bayesian ages in well-studied detached binary systems. Here
we selected $\beta$ Aur to minimise the uncertainties in the bolometric
corrections. The values for the parameters are given by \cite{tag2010}. The
posterior PDF in age for component A is given by the red distribution, while
the green curve gives the PDF for component B. In spite of a secondary
peak in the latter, they have the same mode, and their medians are very
similar. If we impose the constraint that they ought to have the same age,
the prior is then  $\pi(t)=\delta(t_A - t_B)$ resulting in a joint
PDF given by the blue curve, with a much reduced FWHM range.
}
\label{fig:dvg:bin1}
\end{figure}

It is
therefore quite natural to check how their own colour-magnitude diagrams
can test stellar evolution, when the assumption is made that both
components evolved independently of each other (\ie, there were no
mass transfer episodes). 

While there have been many attempts at using classical frequentist
statistical methods \citep[\eg,][and references therein]{erwan2003,young2001,malkov2010}, 
the Bayesian framework presented in the previous section allows one to
infer more robustly the distribution functions of the parameters we are
interested in. While a detailed analysis is beyond the scope of this chapter, 
we can easily check, as an example, the effects of different observables on
the inferred ages of the components. 

\begin{figure}[!ht]
\includegraphics[width=\textwidth]{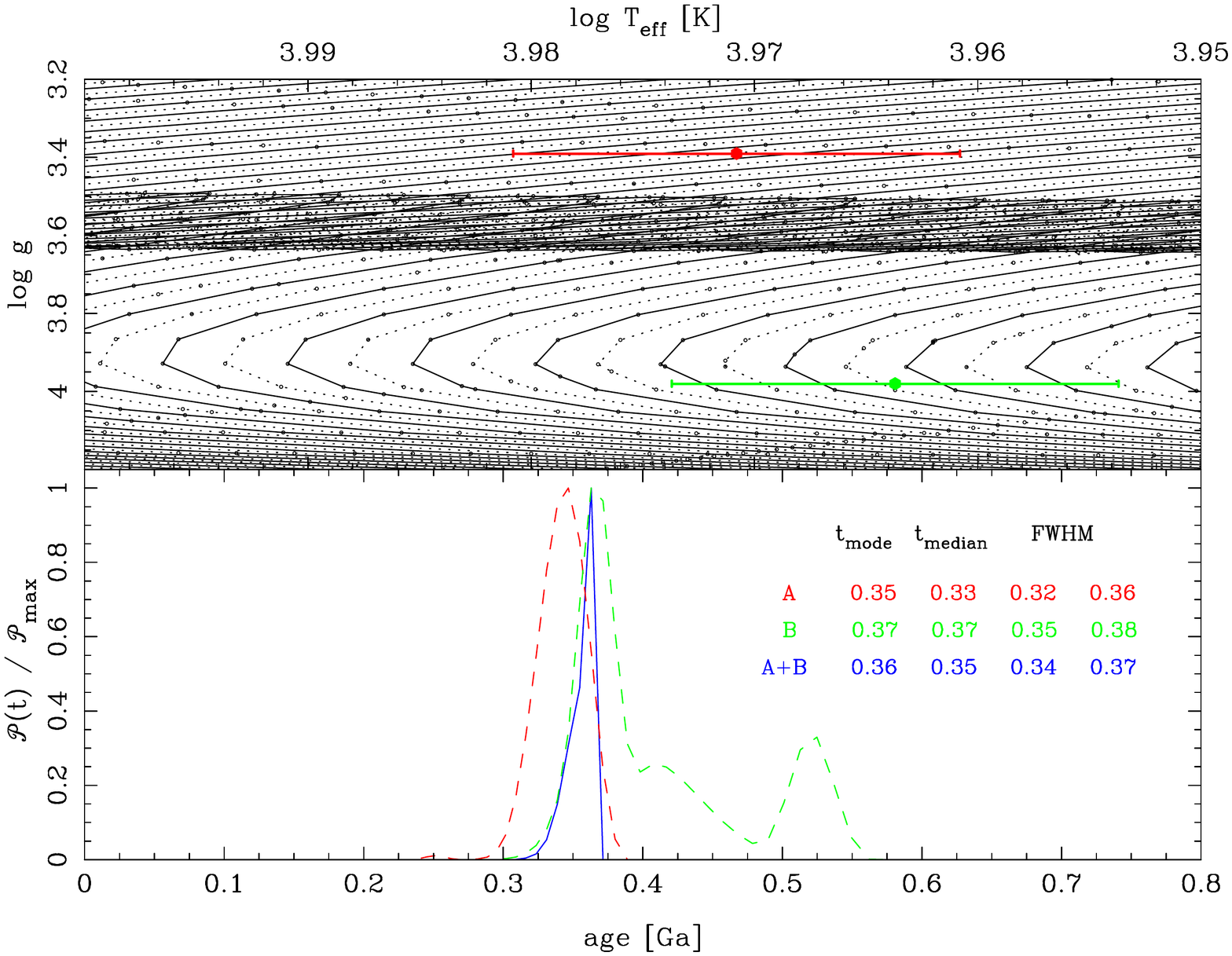}
\caption{The double-lined eclipsing binary $\beta$ Aur (as in Fig.~\ref{fig:dvg:bin1})
this time using effective temperatures and gravities, whose measures are assumed to be
uncorrelated. The error bars  $\sigma(\log g) \sim 0.005$ are too small to be
seen at the scale of this diagram. While the posterior modes are different, the ranges are very similar.
The joint distribution assuming coevality yields a modal age which is significantly 
younger than the one inferred using absolute magnitudes and effective
temperatures (Fig.~\ref{fig:dvg:bin1}).
}
\label{fig:dvg:bin2}
\end{figure}

\begin{figure}[!ht]
\includegraphics[width=\textwidth]{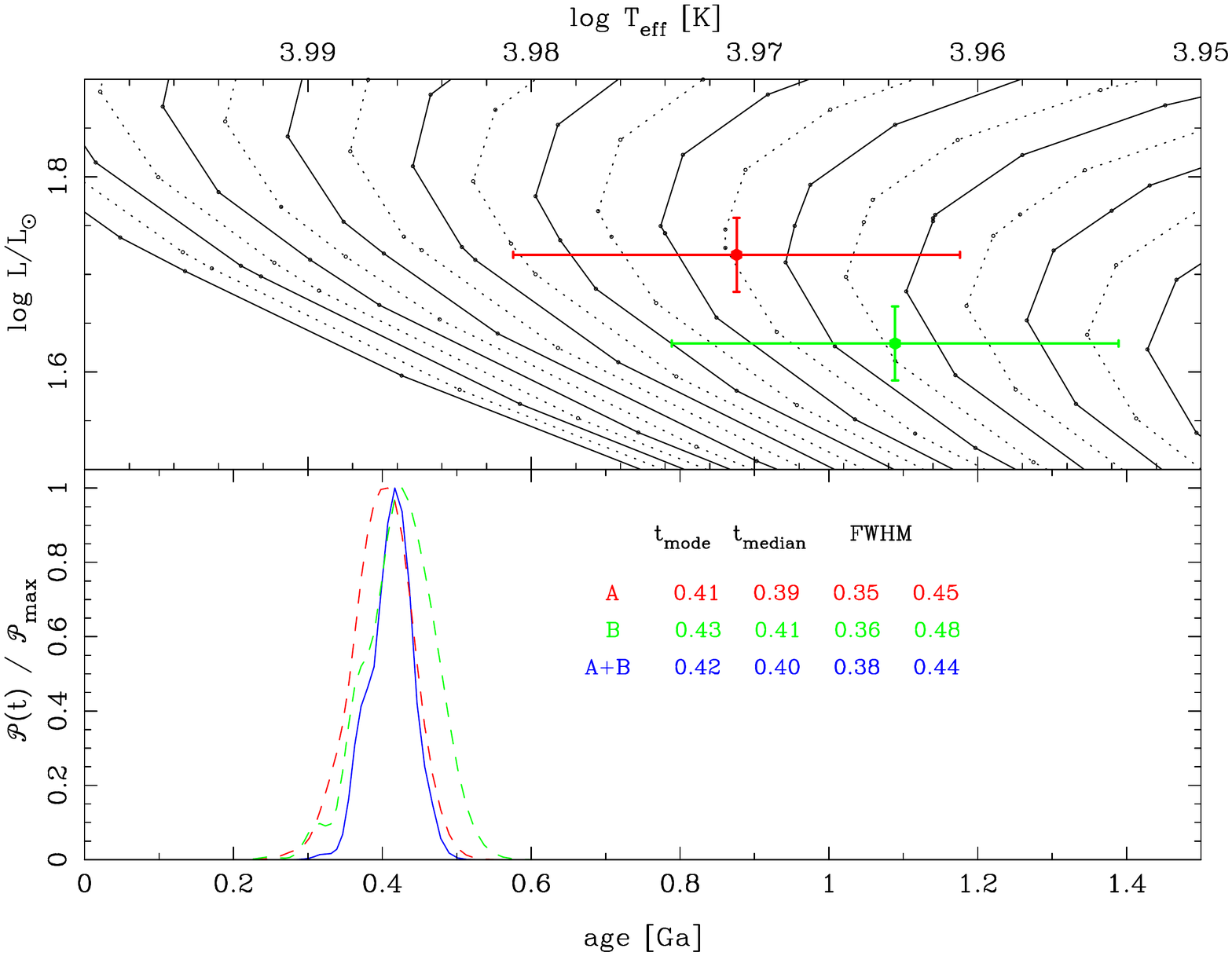}
\caption{Posterior PDF for the age of $\beta$ Aur, using effective
temperatures and absolute luminosities, assumed to be uncorrelated. As
the bolometric corrections at these temperatures are small, it is
not surprising to get results similar to those using absolute
magnitudes (Fig.~\ref{fig:dvg:bin1}), although the distributions 
appear to be wider.
}
\label{fig:dvg:bin3}
\end{figure}

We use the \texttt{BayesGM} formalism to infer the posterior
distribution function of age for each component of the well-studied 
binary system $\beta$ Aur \citep{tag2010}. Fig.~\ref{fig:dvg:bin1} shows
the resulting PDFs, with no prior assumption on their possible coevality. 
Their posterior modes coincide at an age of 0.41 Ga when using effective
temperatures combined with absolute magnitudes in the $V$ band. 
Fig.~\ref{fig:dvg:bin2} shows the same posterior age distributions when 
another pair of observables is used: gravity and effective temperature. In
this case the posterior PDFs are more concentrated, but their modes are
significantly different from the ones using absolute magnitudes and temperatures.
In spite of small bolometric corrections, using absolute luminosities rather
than absolute magnitudes in $V$ yield wider PDFs (Fig.~\ref{fig:dvg:bin3}), 
with ages consistently
larger than the ones inferred from the ($\log g, \log T_{\mathrm{eff}}$) pairs. 
Remarquably, in all cases the PDFs for each component overlap, arguing 
strongly for coevality (but note this was not assumed a priori). If we
impose the condition of coevality, the likelihoods of each component are
combined and multiplied by a Dirac distribution $\delta(t_A - t_B)$. The
resulting PDFs are shown with solid lines in these three figures. As expected in
the Bayesian framework, the more information included results in posterior
distributions which are much narrower.

While there are no other independent constraints on age, one could use the
isochrones to estimate the posterior PDFs on the masses, which are measured
independently, or else use the measures as priors. This can very easily done
in the definition of the physical likelihood.

One can also combine prior information when binaries are known to be
members of a cluster: as their share the same distance, one can impose this
condition in the likelihood to get better constraints on the other
parameters \citep[\eg,][]{lastennet1999}, even though in the case of very 
nearby clusters the depth or extent along the radial direction may be an issue. 

Systematics are likely to become the major source of uncertainties in these
analyses. Limb
darkening `laws' appropriate for the stars in consideration, the amount of
`third light', the sources of noise, etc \citep{southworth2011}. Yet, 
the advent of both massive variability surveys and space missions is providing 
light curves of such quality \citep{bruntt2008} that empirical methods can no
longer be reasonably used.

\section{Coeval stellar populations}
\label{dvg:coeval}

The \textsl{quantitative} comparison of synthetic CMDs with observations relied initially on
either post-MS phases, or the ratio of giants to main-sequence stars, the tips of various
loops, etc \citep[see, \eg,][for some early attempts]{meyer1969,robertson1974,becker1984}.
The advent of high-quality CCD photometry and the increase in computing power made it possible
to apply proper statistical tools to the problem of \textsl{inverting} the observed 
CMDs to infer the underlying physical properties.  \citet{patenaude1978}
 is an example in the attempt at setting up empirical isochrones for open clusters, for
an easier comparison with theoretical ones. 
In fact, 
observers have been fond of establishing the so-called `semi-empirical methods' whereby the
separation between characteristic points in the CMD are calibrated with theoretical
models. Hence we have for instance the vertical separation between the turn-off (TO) and the horizontal
branch, or the horizontal separation between the TO and some point in the sub-giant
branch. These distances are in fact ill-defined, as photometric errors, binaries, and
the intrinsic sampling contribute to a dispersion which is difficult to quantity. Detailed
attempts at calibrating these `empirical' distances \citep[\eg,][]{meissner2006} must be
superseded by a proper modern statistical framework.

While producing synthetic CMDs used to be costly, there are many tools available based
on different set of evolutionary tracks to create isochrones, luminosity functions, 
integrated magnitudes (Table~\ref{dvg:table1}).

\begin{table}[!htb]
\caption{Some popular codes for generating isochrones and CMDs.}
\label{dvg:table1}
{\small 
\begin{tabular}{ll}
\hline \hline
\texttt{CMD v2.6} & \url{stev.oapd.inaf.it/cgi-bin/cmd} \\
Victoria-Regina   & \\
\multicolumn{2}{l}{ \small{%
\url{www2.cadc-ccda.hia-iha.nrc-cnrc.gc.ca/community/VictoriaReginaModels}}}\\
Darmouth          & \url{stellar.dartmouth.edu/models}  \\ 
\texttt{IAC-STAR} & \url{iac-star.iac.es/cmd/index.htm} \\ \hline  
\end{tabular}
}
\end{table}

\citet{naylor2006} proposed a maximum likelihood method whe\-re\-by a simulated underlying
distribution function with some a priori information is made, and then maximise a
geometric likelihood to assess a goodness of fit. The advantage of this approach is
that it allows them to include populations of (unresolved) binaries, where both
the fraction of binaries and the mass ratio distribution can be accounted for.  Their
code is available at the website \url{www.astro.ex.ac.uk/people/timn/tau-squared}.

The cross entropy technique has also been used in solving the optimisation
problem \citep{monteiro2010}, although in this case the authors used a geometrical
likelihood, weighted through Monte Carlo realisations, which, in effect, reduce
to the Bayesian formulation, although in a rather convoluted way. 

\begin{figure}[!ht]
\includegraphics[width=\textwidth]{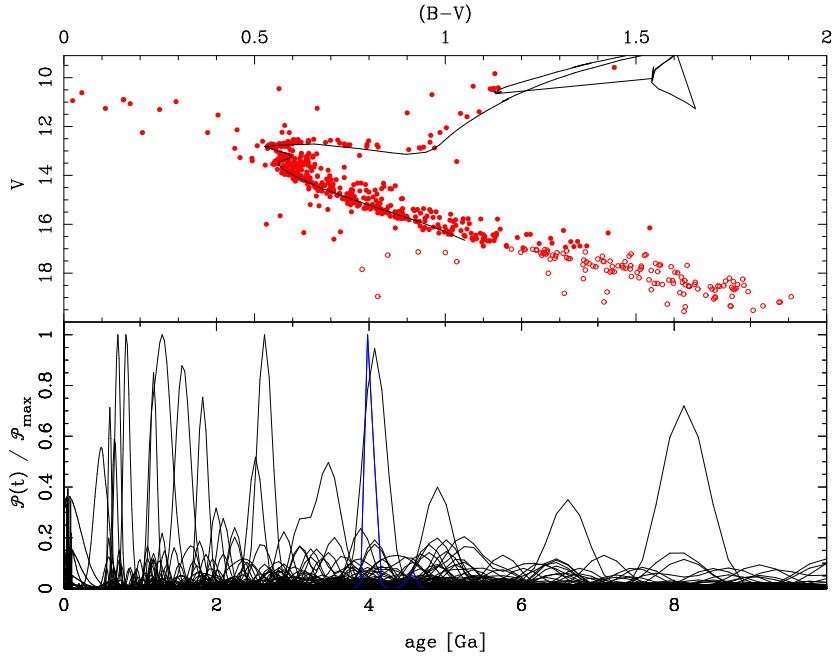}
\caption{\texttt{BayesGM} can be used to determine the ages of clusters in
the same way as in detached binaries, under the assumption of
coevality of its stellar populations, in this case M67 (NGC 2682). Each star produces its age PDF (black
lines in the lower panel), whose
product yields the cluster's age PDF (blue curve). A prior on metallicity
of [Fe/H] = 0.0 $\pm$ 0.1 is assumed, given the spectroscopic
measures of both dwarf and giant stars \citep{santos2009}. Only stars with
probabilities of membership larger than 50\% based on proper motions and
radial velocities have been used \citep{yadav2008}. Only stars brighter than
$V$=17 are considered, as some sets of tracks appear to present bluer
isochrones than observed \citep{yadav2008} at fainter magnitudes. The
mode of the marginalised PDF on distance modulus is $\mu=9.65$.
}
\label{fig:dvg:cum1}
\end{figure}

We can use \texttt{BayesGM} to infer, for example, the age of a coeval stellar population
 just as we did for a binary system. We form the combined likelihood of the 
$\Nstar$ stars with $n$ observables, say, $n=2$ with (V, B-V), as
\beq
\like_{combined}(V,B-V|\btheta) \; = \; \prod\limits_{j=1}^{\Nstar} 
\; \like_{phys}(V_j,(B-V)_j|\btheta) \; ,
\eeq
where $\like_{phys}(\bdata|\btheta)$ is given in Eq.~\ref{dvg:eq:likephys2} 
and we have the equivalent posterior PDF (Eq.~\ref{eq:dvg:post}. Note that, at
this stage, we are not imposing a priori than the stars must be coeval. 

\begin{figure}[!ht]
\includegraphics[width=\textwidth]{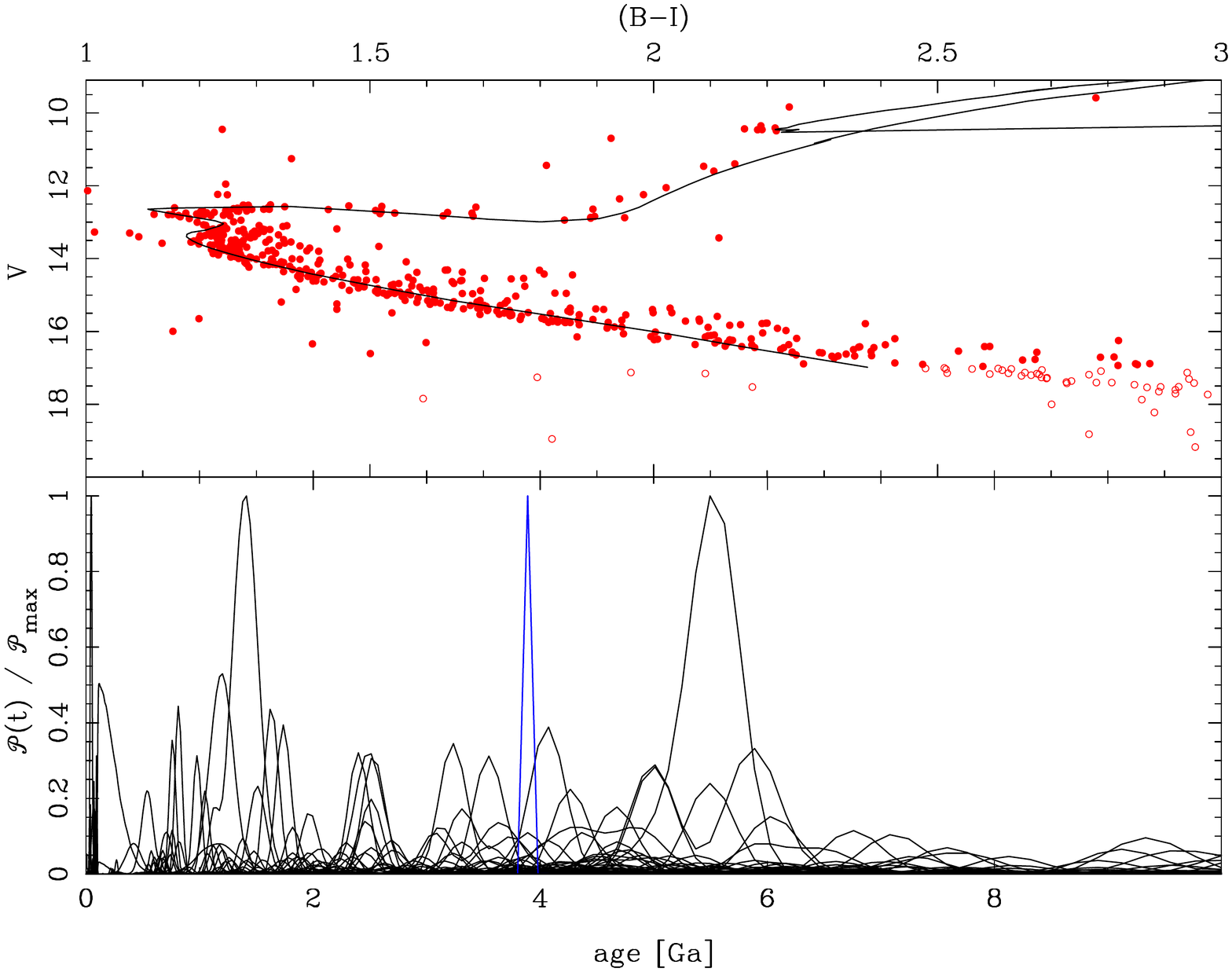}
\caption{The same dataset for M67 as in Fig.~\ref{fig:dvg:cum1}
 but this time using the $B, V$ and $I$ photometry yields the same modal age, but a mode
of the posterior PDF in distance of $\mu = 9.53$. The difference cannot be accounted
for extinction and shows the level of systematics at $\delta\mu \sim 0.1$.}
\label{fig:dvg:cum2}
\end{figure}

The results of the exercise for the M67 open cluster are shown in Fig.~\ref{fig:dvg:cum1} 
where the age PDF has been obtained by marginalising over the other parameters
 (in this case, distance, reddening, metallicity). Each star yields its PDF in the
parameters, and the lower panel shows the full range of PDFs reached by the ensemble
of stars in M67. Clearly some stars cannot possibly be members of the cluster, for
they have widely discrepant age distributions. This technique allows one to make
a further selection (besides proper motions) for true members of the cluster. Note also
that some distributions may be affected by some of the underlying assumptions made: 
some stars are certainly unresolved binaries and clearly the isochrones cannot
possibly fit the blue stragglers present which will, nevertheless, produce unsensical
PDFs unless they are filtered out. 

The product of all the marginalised PDFs yields the PDF of the age of the ensemble 
as $\pi(t)=\delta(t-t_{\star})$ 
and in this case its mode lies at 3.89 Ga, with a sharply-defined PDF. The corresponding
isochrone is indicated in the upper panel of Fig.~\ref{fig:dvg:cum1}. The open circles
in the figure are stars fainter than $V=17$ and which appear to be too blue for the
set of isochrones (this problem is also found in a different context by \citet{an2007}). Could
this be caused by chromospheric activity or is it a problem in the evolutionary tracks?

To check this issue, we can take the set of $(B,V,I)$ independent measures and re-do
the analysis. Fig.~\ref{fig:dvg:cum2} shows that in this case the mode of the posterior
age distribution peaks at 3.98 Ga, and the PDF is fully consistent with the one
inferred from the $(B,V)$ set. However, the modes of the distance modulus PDFs are
significantly different, 9.65 and 9.53, and cannot be accounted for the uncertainties
in the reddening (which was also marginalised out). The outcome of the analysis is
that there is a level of systematics that can only be explored using the fully N-dimensional
PDFs, to assess correlations between the parameters and possible causes of 
inconsistencies.

\begin{figure}[!ht]
\includegraphics[width=\textwidth]{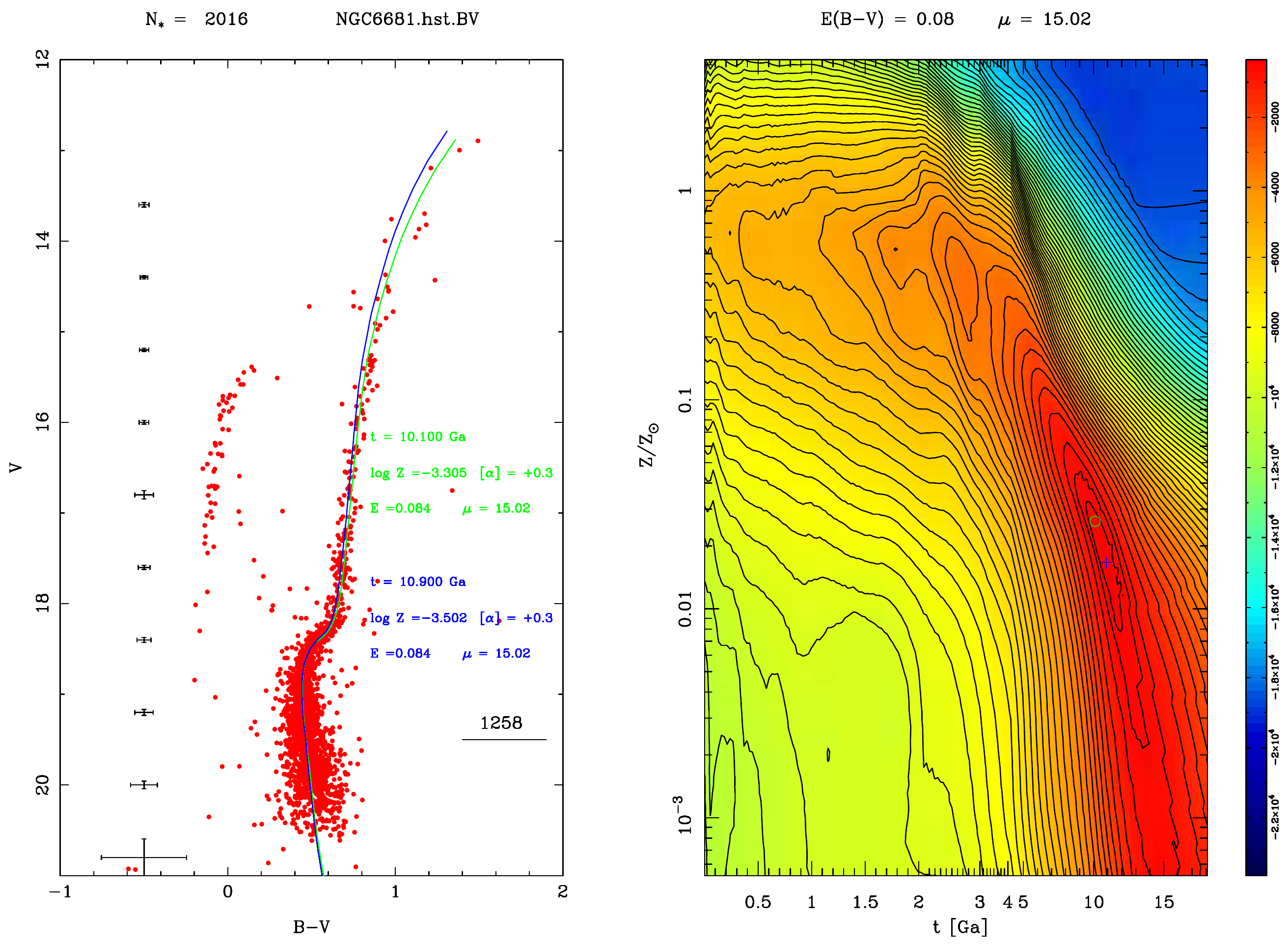}
\caption{Lifting the geometrical age-metallicity degeneracy. The left
panel shows HST observations of NGC 6681, suitably transformed to the UBVRI
system. The typical error bars at each magnitude are indicated on the left. Only the 1258 stars brighter than $V$=19.5 are used. \texttt{BayesGM} yields
a posterior PDF on metallicity of $\log Z = -3.502$ with [$\alpha$/Fe]=+0.3,
a colour excess $E(B-V)=0.084$, distance modulus $\mu=15.02$ and age
$t$=10.9 Ga (blue isochrone, plus sign on the right panel). The solution
depends critically on the possible membership of three stars at the top
of the RGB. Removing these three stars yields the blue isochrone, while
assuming that they are members produces an age which is older by 0.8 Ga,
the remaining posteriors being the same. In contrast with Fig.~\ref{fig:dvg:degenerate}
where the locus of maximum likelihood followed a long stretch, here the
inclusion of the evolutionary speed along each isochrone creates closed
contours (right panel), and hence lifts the degeneracy.
}
\label{fig:dvg:6681}
\end{figure}

We can now see an example of the multi-dimensional PDFs case by applying \texttt{BayesGM} to
globular clusters, with flat priors\footnote{Note however that the 
set of stellar tracks used 
only allowed three different values for [$\alpha$/Fe], namely +0.0, +0.2 and +0.3.}. 
Fig.~\ref{fig:dvg:6681} shows the CMD of the
old globular cluster NGC 6681, observed with HST, assuming the standard transformations
to the $B,V$ system. Here the modes of the posterior appear well-defined as well, and
the right panel of the Figure shows the PDF marginalised over all parameters but 
age and metallicity. The probability contours have the same shape as the ones we saw
in the age-metallicity degeneracy (Fig.~\ref{fig:dvg:degenerate}), except that this
time we can lift entirely the degeneracy: only a tiny area has the maximum 
probability.

The technique is very powerful, yet subject to some interesting systematics. 
Fig.~\ref{fig:dvg:6681} also shows the resulting isochrone for the modal age (and
metallicity, distance, etc) when the top 3 stars are removed. These are very
bright stars and one may wonder whether they do belong to the cluster at all. In
this case, the modal age shifts by 0.8 Ga to younger ages (still within the top
most inner probability contour), and the modal posterior metallicity moves 
-0.2 dex. This is not, however, a sign of degeneracy because the data set is
different, it just shows the sensitivity of some parameters to outliers (the
distance and reddening are, quite rightly, unaffected by their presence).

\begin{figure}[!ht]
\includegraphics[width=\textwidth]{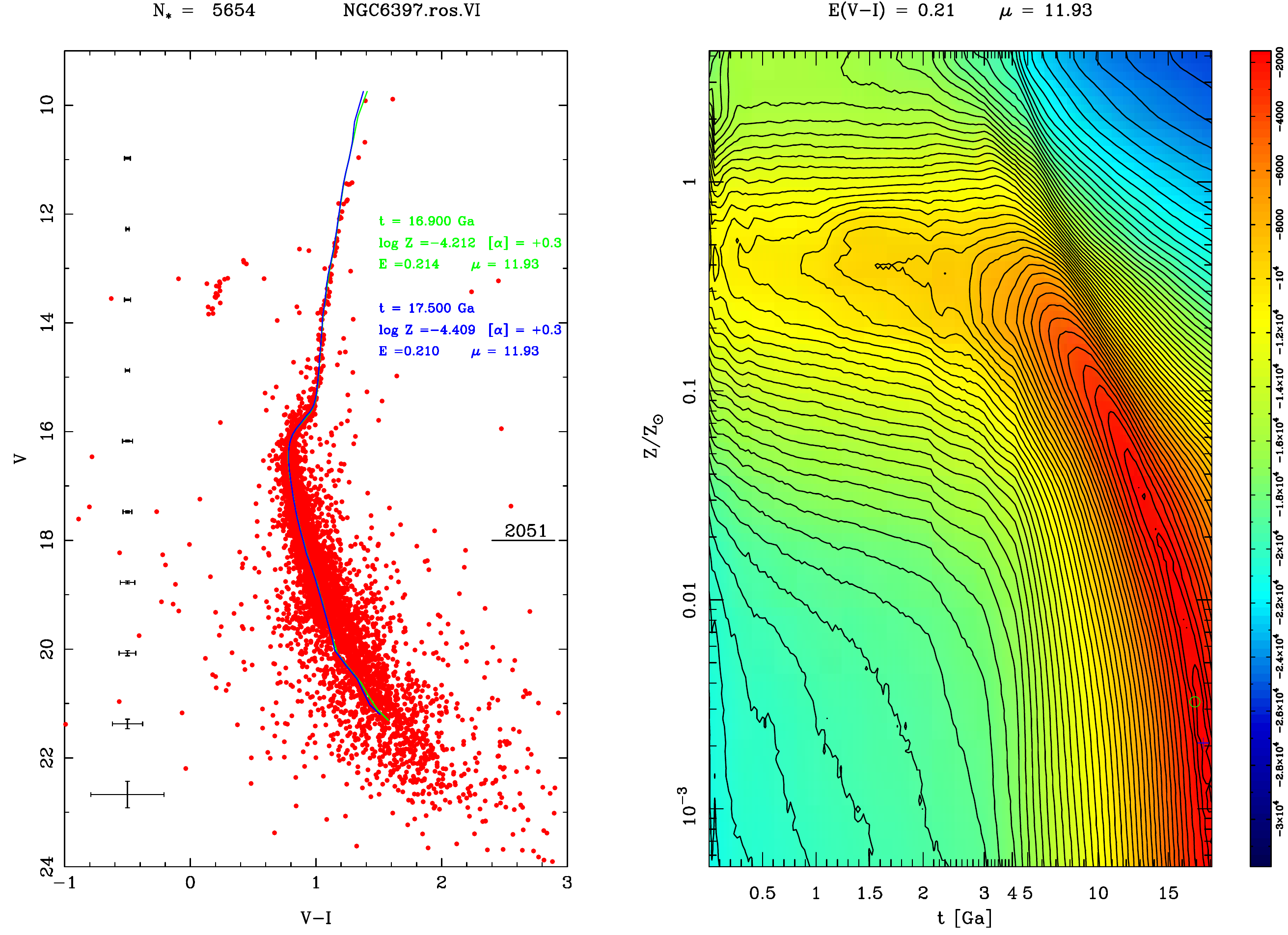}
\caption{Ground-based observations of NGC 6397 yield 2051 stars brighter
than $V$=18, with average photometric uncertainties as indicated on the left.
The age-metallicity degeneracy is lifted (right panel), although in this
case the issue of whether the four topmost stars in the RGB belong to the
cluster or not does not change the mode of the posterior PDF of the age:
 in both instances the ages appear to be larger than 16.5 Ga, which may
point to a limitation in the stellar evolutionary tracks at these low
metallicities.
}
\label{fig:dvg:6397}
\end{figure}

Outliers and possible non members do not always perturb the modes. 
Fig.~\ref{fig:dvg:6397} shows the ground-based CMD of NGC 6397 and a sharply-peaked 
marginalised 2-dimensional PDF with modes at 16.9 Ga and $Z=-4.2$ dex. In this case, 
removing the four brightest stars shifts the modes to 17.5 Ga and $Z=-4.41$ dex. Taken
at face value, this globular cluster appears older than the age of the universe
as inferred from a set of completely independent measures (CMB fluctuations and
the expansion rate), but in fact points to a systematic in this set of tracks
at these very low metallicities\footnote{Since the dataset has been modified 
as some stars has been excluded, note that the evidence changes, and hence the
posteriors must be compared properly normalised.}.

An important caveat is that an increasingly large number of mainly spectroscopic
(but also some photometric) observations is 
revealing that many globular clusters have multiple, not simple stellar populations.
The spread and anti-correlation of Na and O, for example, may be accounted for
 in a self-pollution scenario, where the ejecta from the old population lead to
a composition of the younger one enriched in He, N, Na, Al, but depleted in C, O, Ne
and Mg. However, the Bayesian technique can, in fact, assign individual
probabilities of membership to one or another of the population, precisely because
their different abundances may lead to differential evolution and hence positions
in the CMDs.

\section{Composite resolved  stellar populations}
\label{dvg:composite}
A variable star formation rate and chemical evolution history give rise to a
composite stellar population, which is a mixture of stars of different ages and chemical
abundances. Disentangling which populations were formed in this scenario is the
main goal of the general inverse problem: \textsl{Given an observed CMD, what is the distribution of ages of its stellar populations?}. This distribution
is the basis with which inferences on the 
star formation and chemical enrichment histories can be made.  

The problem is far from trivial, in part due to the apparent age-metallicity 
degeneracy, and was limited during a long time to a qualitative comparison between
the observations and synthetic CMDs with prescribed SFR and $Z(t)$ histories. 
\citet{aparicio1990} and \citet{tosi1991} were the early pioneers to attempt the
statistical inversion of the CMDs based on the comparison of number counts in 
suitably-defined areas of 
the CMD, and were a step beyond a qualitative analysis. There soon were many other
attempts \citep[see, \eg,][for a review]{gallart2005} in this direct approach, 
including \citet{ng2002} who used an optimisation technique with 
a genetic algorithm. The
thorny issue, as we will see, is to compare properly simulated CMDs with observations
(Section \S~\ref{dvg:stats}).

\citet{dolphin1997} (later refined and applied in \citet{dolphin2002}'s  
\texttt{MATCH} code at \url{americano.dolphinsim.com/match}), \citet{olsen1999} and
\citet{harris2001} (\texttt{StarFish} code \url{www.noao.edu/staff/jharris/SFH}) proposed
to decompose a generic CMD into a linear
combination of `elemental' CMDs produced by coeval populations of well-defined
ages and metallicities. This is also the method use by 
\citet{makarov2004} with their \texttt{StarProbe} code. The observed CMD is thus posited
to be 
produced by the linear combination of `partial' CMDs as
\beq
N(i,j) \; = \; \sum_k \; R_k \; n(i,j)_k \quad ,
\label{eq:dvg:partial}
\eeq
where $N(i,j)$ is the number of stars in the bin ($i,j$) of the observed CMD, made up
of the weighted sum of $k$  partial CMDs with counts $ n(i,j)_k$ in that bin. If the
partial CMDs were computed for a nominal $SFR = 1 \Msun\yr^{-1}$, then the weights
$R_k$ provide the star formation rate contributed by the $k$-th partial CMD. One can add a
foreground CMD to model any underlying contamination, completeness, etc. 

Different statistics have been used to infer the values of $R_k$ for the
set of partial CMDs, although none of them is entirely correct (\cf \S~\ref{dvg:stats}).
Imposing the non-negativity constraint that $R_k \geq 0 \; (\forall k)$ allows one
to determine them through iterative or steepest descent methods \citep[in some cases 
claims of a unique solution have been made, see][]{dolphin2002}. For a grid of,
say, $10\times 10$ CMDs spanning 10 ages and 10 metallicities, this is a problem
of finding the absolute minimum in a parameter space with $K=100$ dimensions, a non
trivial task given the likely presence of many secondary minima. For this reason, 
 the parameter
space can be efficiently explored using genetic algorithms to find the absolute maximum 
corresponding to the best fit \citep{ng2002,aparicio2009}, as many secondary minima do 
exist in this highly dimensional optimisation problem. Their code, \texttt{IAC-POP},
 is available at: \url{www.iac.es/galeria/aaj/iac-pop_eng.htm}. 

\citet{tolstoysaha1996} pioneered a Bayesian formulation of the problem by pondering
on the method to compare datasets drawn from simulations with the actual observed
CMD, while \cite{cignonishore2006} used the Richardson-Lucy technique to deconvolve 
the observed CMD in order to produce a `reconstructed' CMD which can then be
compared with simulations.

In fact, the proper way to formulate the problem is to realise that this is, in
statistical parlance, an inverse problem which is \textsl{ill-defined} and may
have multiple solutions \citep{craigbrown1986}. Yet claims have
been made that unique solutions can nevertheless be found, not only to the functionals
defining the star formation and chemical enrichment histories, but also the shape
of the IMF, the mass ratio distribution of the unresolved 
binaries, etc \citep{vergely2002,wilson2003}. To some extent, this may perhaps 
be true, but
as in all ill-posed this comes at the prize of a compromise between accuracy/smoothness
 and resolution. To see this, write the probability that a given star with dataset
$\bdata_i$ comes from a star formation episode at age $t$ when the star formation
rate was $SFR(t)$ and the metallicity $Z(t)$:
\begin{eqnarray}
\mathcal{P}\left[SFR(t),Z(t)|\bdata_i\right] \; & = &  \nonumber \\
\; \pi(t) \; \pi(Z) \; 
\int\limits_{t_0}^{t_1} \int\limits_{Z_0}^{Z_1} \; SFR(t) \; 
\like_{phys}\left(\bdata_i|\btheta,Z(t)\right) \; dt \; dZ \; &  \;& ,
\end{eqnarray}
where the limits of the integrals come from our prior knowledge $\pi(t)$ 
that the star must
have an age between these limits (could be a least informative prior between
 0 to 15 Ga), and $\pi(Z)$ describes
our prior probability on metallicity (could be flat between $-5$ and 1, or limited
to some range if we have previous measures of $Z$ for these stars).
 For an ensemble of $\Nstar$ stars
the combined probability that their full CMD arises from the episodes of star
formation described by $SFR(t)$ and chemical enrichment given by $Z(t)$ becomes
\beq
\mathcal{P}_{CMD}\left[SFR(t),Z(t)|\bdata\right] \; = \; \prod\limits_{i=1}^{\Nstar} \; 
 \mathcal{P}\left[SFR(t),Z(t)|\bdata_i\right] \quad .
\eeq
Clearly, given some precision in the photometric data, the fine-grained details
of the functions $SFR(t)$ and $Z(t)$ are impossible to constrain, hence the
infinity of possible solutions, and the ill-posedness nature of the problem.
To regularise the problem one seeks not to maximise this probability, but rather
the log of the probability to which one adds a regularisation term which depends
on the type of solution one seeks \citep{craigbrown1986}. The choice varies between terms which penalise
variations (\ie, imposes constant solutions to the functions), or terms which
penalise gradients or large deviations (\ie, imposes a more or less
strict smoothness to the solutions).

The general problem, as far as we are aware, has not yet been solved, and only
partial solutions have been explored. For instance, in the case where our prior
in metallicity is peaked at some well-defined value $Z_{\star}$ (as is the case in some
dwarf galaxies, whose dispersion in metallicity appears to be quite small), one
can assume that $\pi(Z) = \delta(Z-Z_{\star})$, and the problem reduces to solving
for only one function, $SFR(t)$. One could impose a parametric form to this function
(say in terms of a piece-wise constant function, or a series of Gaussian bursts) and
solve for the parameters of the assumed function. Instead, \citet{Hernandez:1999lo}
realised that seeking the maximum of the probability function is equivalent to
ask that the variation of the function is zero, ensuring that it is an extremum. 
In turn, using the Euler-Lagrange equation, 
$\delta\mathcal{P}_{CMD}[SFR(t),Z_{\star}]=0$ implies a set of $\Nstar$ coupled
differential equations that can be easily solved. In this way a fully non-parametric
solution of $SFR(t)$ was built, with no prior information on shape, form or amplitude.
The technique was applied to some nearby dwarf galaxies \citep{Hernandez:2000hs}, 
whose HST-based CMDs were
well-measured\footnote{\citet{dolphin2002} rightly noted that some zero point offsets which were
adopted were wrong, shifting the solutions by a small amount. Much work
remains to be done on systematics.}, as well as to
the solar neighbourhood probed by Hipparcos \citep{Hernandez:2000et} with a wealth of detail in spite
of the few stars used. This was also explored independently by \citet{cignoni2006} 
with an entirely different method, and there are many more results in both
clusters and nearby galaxies using a variety of methods, most of which
are described in  recent review articles 
\citep{tolstoy2009,cignonitosi2010}. 

Interestingly, even if the general problem of determining \textsl{both} functions
independently can be solved, we know from chemical evolution theory that there
must be some coupling: an enhanced episode of star formation will lead to an
enrichment on some time scale which depends on the element and on the details
of the nucleosynthetic yields, among other things. 
\citet{yuklee2007} propose to compute in a self-consistent way the chemical
enrichment history while inferring at the same time the star formation rate 
history. However, this is clearly some prior information that is imposed, and
it is far from clear if, for instance, a closed box model or the assumed recycling 
are correct. At some point, one can imagine the infall of unenriched gas, or
else the accretion of a satellite bringing enriched material. 
There is clearly ample scope for progress in this field. 
 
\section{Unresolved nopulations and pixel CMDs}
\label{dvg:pixels}
The extreme case of composite populations is reached when their stars cannot be
resolved. This is the case, for instance, when analysing the integrated spectra
of galaxies or clusters: we only have access to luminosity-weighted estimates of
the quantities of interest (age, $Z$, etc). Given the shape of the luminosity function, where
a handful of bright stars can dominate the flux of the population (which  in
fact is dominated in number and mass by the less massive stars), one can easily
see the major biases inherent to these techniques.

An intermediate case, which is essential to understand these stellar populations,
is the one in the pixels of images of nearby galaxies. In this case, we no longer have to
deal with billions of stars (integrated spectra) but rather some $10^2 \cdots 10^4$
 stars or
so, depending of course on the distance and the size of the pixels. A formalism
was  proposed \citet{renzini1998} and observations were pioneered by \citet{bothum1986}
with ground-based observations. 
 \citet{abraham1999} discussed, in a forward modelling approach, the
interpretation of the 4-band images of galaxies in the Hubble Deep field, and the
limitations produced by the extinction-colour-metallicity degeneracy.  
Not limited by the seeing, HST-based studies have dominated the field, which is
becoming an essential tool for the understanding of galaxy evolution 
\citep[e.g][]{conti2003,kassin2003,eskridge2003,lfcm2007,lee2011}. 

While a proper formulation of the inverse problem is still lacking, 
the forward modelling techniques 
--which has so far been used-- assume a full sampling of the underlyng stellar 
populations.
This may be appropriate for integrated properties, but not in pixels, where the
number of stars, while important, is not large enough to ensure the
statistical convergence in the
properties. This is also the case in the r\'egime where the star formation rate is
low, and, in general, in places where the number of stars is small enough as to
create stochastic variations in the properties such as luminosities and colours. The
importance of these fluctuations is essential also to assess whether the IMF
is universal: could the stochastic variations be consistent with samples drawn from
the same IMF ?

This stochasticity is equivalent to a lack of convergence in the properties and 
can be quantified in a simple way, either assuming
quasi-Poisson counts \citep[\eg,][]{Cervino:2003sj, Cervino:2002yb}, and through the
concept of the lowest luminosity limit, a limit which ensures that
statistical fluctuations become unimportant \citep{cervino2006, Cervino:2009sp}. 
Another approach, more costly in CPU time, is to create Monte Carlo samples. 
For example \citet{popescu2009} proposed a code, \texttt{MASSCLEAN} (available
at \url{www.physics.uc.edu/~bogdan/massclean.html}) to carry out multi-colour
 simulations which do not assume a full sampling of the IMF, and confirm by and
large the analytical predictions in the quasi-Poisson r\'egime discussed above. This
approach is likely to change entirely the interpretation of the integrated
properties of stellar clusters \citep{popescu2010a,popescu2010b,popescu2012}. In a
similar way, new tools have been put forward to produce these stochastic variations in
stellar populations. \citet{dasilva2012} present a code, \texttt{SLUG} 
(\url{sites.google.com/site/runslug}) which is also likely to revisit many results
obtained thus far in  galaxy evolution.

\section{Best-fit solutions and uncertainties}
\label{dvg:stats}
An issue which is seldom addressed, if at all, is whether the best-fit solution found 
(by any method) is also a good fit. There is no guarantee whatsoever that the
best straight-line fit to a parabola is a good fit, even if it is the best of
all possible within the assumption of a straight line.
In some cases the problem is not even tackled. Sometimes a $\chi^2$ criterion 
is used (which is not
appropriate, since number counts in any CMD cell or bin follow Poisson statistics), or a comparison with
a \textsl{single} model realisation is performed (for instance analysing the 
residuals in a Hess diagram,
which is just a binned version of the CMD, as first used by \citet{hess1924} in a
different context).
None of these approaches is satisfactory: one has to deal \textsl{both} with 
Poisson counts and with
the intrinsic variability in the \textsl{model} predictions (as different 
realisations of the model will
invariably yield different results and hence residuals).

Some statistics that have been used in the context of comparing model CMDs 
(with $\{m_i\}$ stars) 
with observed CMDs (with $\{s_i\}$ stars) in a selection of $B$ bins or
boxes  include the following:
\begin{enumerate}
\item Pearson's $\chi_P^2$. This takes into account the variability of the
model counts $\{m_i\}$ only:
\beq
\chi_P^2 \; = \; \sum_i^B \frac{\left( s_i - m_i \right)^2}{m_i} \; .
\eeq
\item Modified Neyman's $\chi_N^2$ which adopts the form
\beq
\chi_N^2 \; = \; \sum_i^B \frac{\left( s_i - m_i \right)^2}{\mathrm{max}(s_i,1)} \; ,
\eeq
but only encapsulates the variability of the observed counts $\{s_i\}$.
\item Dispersion. \cite{kerber2001} minimise the dispersion defined as
\beq
\mathcal{S}^2 \; = \; \sum_i^B (m_i - s_i)^2 \; .
\eeq
While clearly it has the correct behaviour, it does not take into account
the Poisson distribution of both model and observed counts. Could the minimum
$\mathcal{S}^2$ value reached be different should a different realisation of the model
$\{m_i\}$ be used for the comparison?
\item Percentile position. \cite{kerber2001} also use the percentile
position of $\mid s_i - m_i \mid$ within the distribution of
$\mid m_i = m_{ij} \mid$ with $j=1, \cdots N_{sim}$ and $m_{ij}$ being
Poisson realisations with parameter $m_i$ in each bin $i$. Then the
statistic
\beq
\mathrm{pss} \; = \; \sum_i^B \, \left( 1 - p_i \right)
\eeq
is minimised. This tackles the part of the intrinsic variability of the
model outcomes, but not the one from the observations.
\item $\chi^2$ statistic for Poisson variables. Both \citet{ng1998} and 
\citet{mighell1999} point out that the number counts in the bins the CMD has
been divided into follow Poisson statistics, not Gaussian ones (unless 
we are in the large numbers limit, which will not be the case in
sparsely-populated bins). The statistic proposed is
\beq
\chi^2_{\gamma} \; = \; 2 \sum_i \frac{ \left[ s_i + \mathrm{min}(n_i,1) - m_i \right]^2}{s_i+1} \; \; ,
\eeq
and many codes have implemented this to assess the goodness (or otherwise) 
of their fits. However, comparing Poisson counts is tricky, as all 
high energy physicits know. While the sum of Poisson variates is also
Poisson distributed \citep[see, \eg,][for some consequences in the
context of stellar populations]{Cervino:2003sj}, the difference of
Poisson counts, such as $m_i - s_i$ which one could naively use, do not
follow Poisson statistics but are distributed following a Skellam 
distribution (the fact that the variate may become negative is a clear hint).
\item Poisson likelihood ratio. \citet{dolphin2002} correctly argued that the
proper analogy with the $\chi^2$ ratio (which only applies to Gaussian
statistics) for Poisson variates is the ratio
\beq
- \, 2 \, \ln \mathrm{PLR} \; = 
\; 2 \sum\limits_{i} \left( m_i - s_i 
+ s_i \ln\left(\frac{s_i}{m_i}\right) \right) \; ,
\eeq
as proposed by \citet{bakercousins84}. However, using a likelihood ratio has some constraints that makes this quantity
unsuitable for seeking a proper comparison (see below).
\end{enumerate}

The likelihood of a model is obviously a \textsl{relative} probability. If we
have two models $A$ and $B$ giving likelihoods 
$\cal{L}_A$ and $\cal{L}_B$, model $A$ is $\cal{L}_A / \cal{L}_B$ times as
likely as model $B$. This is also known as the bookmakers' odds
\citep{syersaha1994}. To compare two likelihoods or use a likelihood
ratio, however, two
conditions are essential \citep{protassov2002}, but are too often overlooked:
\begin{enumerate}
\item[\textsl{(i)}] The models must be nested.
\item[\textsl{(ii)}] The values of the parameters must not reach zero.
\end{enumerate}
The first condition implies that, for example, one cannot use a likelihood
ratio for comparing a fit using a polynomial and another one using an
exponential function. It can be used  to compare the fits obtained by
a polynomial of degree $k$ with another fit using a polynomial of degree $j \not= k$.
The second condition, of strict positivity, implies that one cannot use a 
likelihood ratio when decomposing a CMD into a linear combination of partial CMDs, as
quite a few odd partial CMDs are unlikely to contribute at all (say young populations
to a globular cluster).  Necessarily some $R_k=0$ will happen in Eq.~\ref{eq:dvg:partial} 
and the likelihood ratio cannot be applied.

At any rate, $\chi^2$-like statistics such as the ones above 
depend on the size of the bins the CMD is divided into, and care
must be taken when assessing their significance in comparison with CMDs
 with different cell/bin
sizes. In addition, one has to take into account the \textsl{intrinsic} variability of the
model predictions, so a comparison between a \textsl{single} realisation and an 
observed CMD makes very 
little sense. Obviously one can only take into account a Poisson dispersion in
the observed counts, but models have not only this intrinsic variability but
also one associated with them, even for a fixed set of values for the
parameters. 

 A nearly size-independent statistic was suggested
by \cite{bell2008} where the rms deviation of the data with respect to the model is minimised, and
takes explicitely into account the Poisson variability of the model. For a set of $B$ bins, one forms
the \textsl{distribution} of 
\beq
\sigma / \mathrm{total} \; = \; \sqrt{  < \sigma^2 > } \; \left[ \frac{1}{B} \sum_i^B s_i \right ]^{-1} \; ,
\eeq
where
\beq
 < \sigma^2 > \; = \; \frac{1}{B} \, \left [ \sum_i^B (s_i - m_i)^2 - \sum_i^B (m'_i - m_i)^2 \right ] \; .
\eeq
This statistic takes into account the variability of the model, as $\{m'_i\}$ are Poisson realisations
of the model with expectation values  $\{m_i\}$, but \textsl{not} the uncertainties in the observed
counts $\{s_i\}$. It is, in fact, \textsl{designed} to detect the fluctuations in the observed
number counts with respect to the models.

Should we know the distribution function of the model, we could apply the standard statistical
tools to infer the proper credibility intervals. This is obviously not the case in CMDs, but we
can generate samples from the model (a sample of infinite size would be as good as the model). 
Different samples (different realisations) are likely to populate regions of the CMD which may
not coincide with the observed ones, we hence need to smooth out the model and data for a
proper comparison. The simplest way of doing this is binning, with the underlying
assumption that the model distribution function is constant within the bin (hence small bins
are preferred, but they should be large enough that bins contain both 
model and observed stars as much as possible).

If we divide the CMD into $B$ cells,
of arbitrary sizes and shapes,  each containing
$\{s_i\}$ observed stars (with a total of $S = \sum_i^B s_i$ stars),
and 
$\{m_i\}$ model stars (with a total of $M = \sum_i^B m_i$ stars), each cell has a
probability $w_i$ of having the appropriate number of stars (model or observed), 
constant within the bin. The probability distribution function for the bin
occupancies, 
multi-variate distribution of counts, will be given by a
multinomial distribution
\beq
 P(s_i,m_i\mid w_i) = M!\,S!\boxes {w_i^{m_i+s_i}\over m_i!s_i!} \; ,
\eeq
where the weigths  $\{w_i\}$ of the distribution function are unknown. We
only have the constraint, given by the normalisation, that $\sum_i^B w_i = 1$.
We can treat the weights as nuisance parameters and marginalise over
them using the identity
\beq
 \left(\boxes\int\! w_i^{n_i}\,dw_i\right) \, \delta\left(\sum
   w_j-1\right) = {1\over (N+B-1)!} \boxes n_i!  \; ,
\eeq
and we get
\beq
 P(s_i,m_i) = {M!\,S!\,(B-1)!\over (M+S+B-1)!} \zapb  \; ,
\eeq
which is equal to $P(s_i\mid m_i)\, P(m_i)$ .
 And likewise with $P(m_i)$, and so Bayes' theorem gives us
\beq
 P(s_i\mid m_i) = {S!\,(M+B-1)!\over (M+S+B-1)!} \zapb  ; .
\eeq
For a fixed number of bins $B$, model stars $M$ and observed stars $S$, the
first term is a constant and hence 
\beq
{\rm Prob} \propto W = \zapb  \; . \label{eq:dvg:wstat1}
\eeq
This statistic was first proposed by \cite{saha1998} and has been widely used,
for example in comparisons of N-body simulations with discrete data sets 
\citep{sevenster,beaulieu} or indeed
in inversions of CMDs \citep{Hernandez:2000hs,Hernandez:2000et,kerber2002}, 
correlations between the SFR history and the glaciation epochs \citep{delafuente}, 
and in setting constraints on
the properties of clusters \citep[\eg,][]{bruzual,kerber2005}. Contrary to some 
baseless claims
\citep{dolphin2002}, it \textsl{does} allow a proper comparison between data and models.
Note that it takes model counts and observed counts on the same
footing. If $M$ is arbitrarily large, the bins can be made small enough as to contain
one observed star at most, so that the probability goes as $\prod_k (1+m_k)$ where
$k$ is the running index for boxes with $s_k=1$, and with $m_k \gg 1$ we get
the same result as in the continuous distribution case.

More properly in the context of multidimensional Poisson counts, if $M$ and $S$
are not fixed but are the expectations values of the totals when $s_i$ and $m_i$
are drawn from a Poisson process, we get a modified $W$ statistic 
\citep{saha2003} as
\beq
 P(s_i|m_i) = {e^{-M} \, M^M \, e^{-S} \, S^S \,(B-1)!\over (M+S+B-1)!} \zapb \; .
\label{eq:dvg:wstat2}
\eeq
It is clear that, normalisations aside (which can be fixed if indeed
$B$, $M$ and $S$ are kept fixed), Saha's $W$ statistic (Eq.~\ref{eq:dvg:wstat2}) can
be used to perform either
\begin{itemize}
\item \textsl{Parameter fitting}: Just compute the distribution of $W$ for
\textsl{a given data set} and 
different model parameters. That is, fix
$\{s_i\}$, vary model parameters (hence $\{m_i\}$) then read off parameter estimates and 
confidence intervals for the parameters.

\item \textsl{Goodness of fit}: Here we want the distribution of $W$ for \textsl{fixed model parameters}
and various simulated data sets. Fix model parameters (so
$\{m_i\}$ are fixed), then
vary simulated $\{s_i=m'_i\}$. Then compare the \textsl{distribution} of $W$ with the
\textsl{distribution} obtained from the actual data. The extent to which both distributions 
(actually, samples of $W$) can be drawn from the same underlying (and unknown) distribution 
function gives a proper measure of the goodness of the fit.
\end{itemize}

It is therefore the statistic of choice to be used in the context 
of CMD modeling, and turns out to be far superior to the classical
Mann-Withney test for 2 samples. 

In summary, the determination of stellar ages using stellar evolutionary
theory has a proper Bayesian formalism, which allows one to infer not
a single value (which makes little sense from a statistical point of view) 
but rather the posterior distribution function which can be used to
constrain both the properties of the stars and the predictions of the
models. It is beyond the scope of this chapter to analyse the
comparison between different sets of tracks and different assumptions 
on the systematics. This ought to be tackled, in fact, by a double
blind experiment in which a referee generates a set of simulated
CMDs from a given (but unknown to the referee) set of 
isochrones, including realistic photometric errors, 
which are then analysed by different
teams using different isochrones and assumptions. Only in this way
the actual distribution function of ages can be quantitatively 
estimated, and the robust confirgurations which give a narrow
distribution, independently of sets and assumptions, identified.

\section*{Acknowledgements}
The author would like to thank the ``Formation permanente du CNRS'' for
 financial support. This research has made use of NASA's Astrophysics 
Data System.


\vspace*{3.0cm}
\begin{center}
\includegraphics[width=0.8\textwidth]{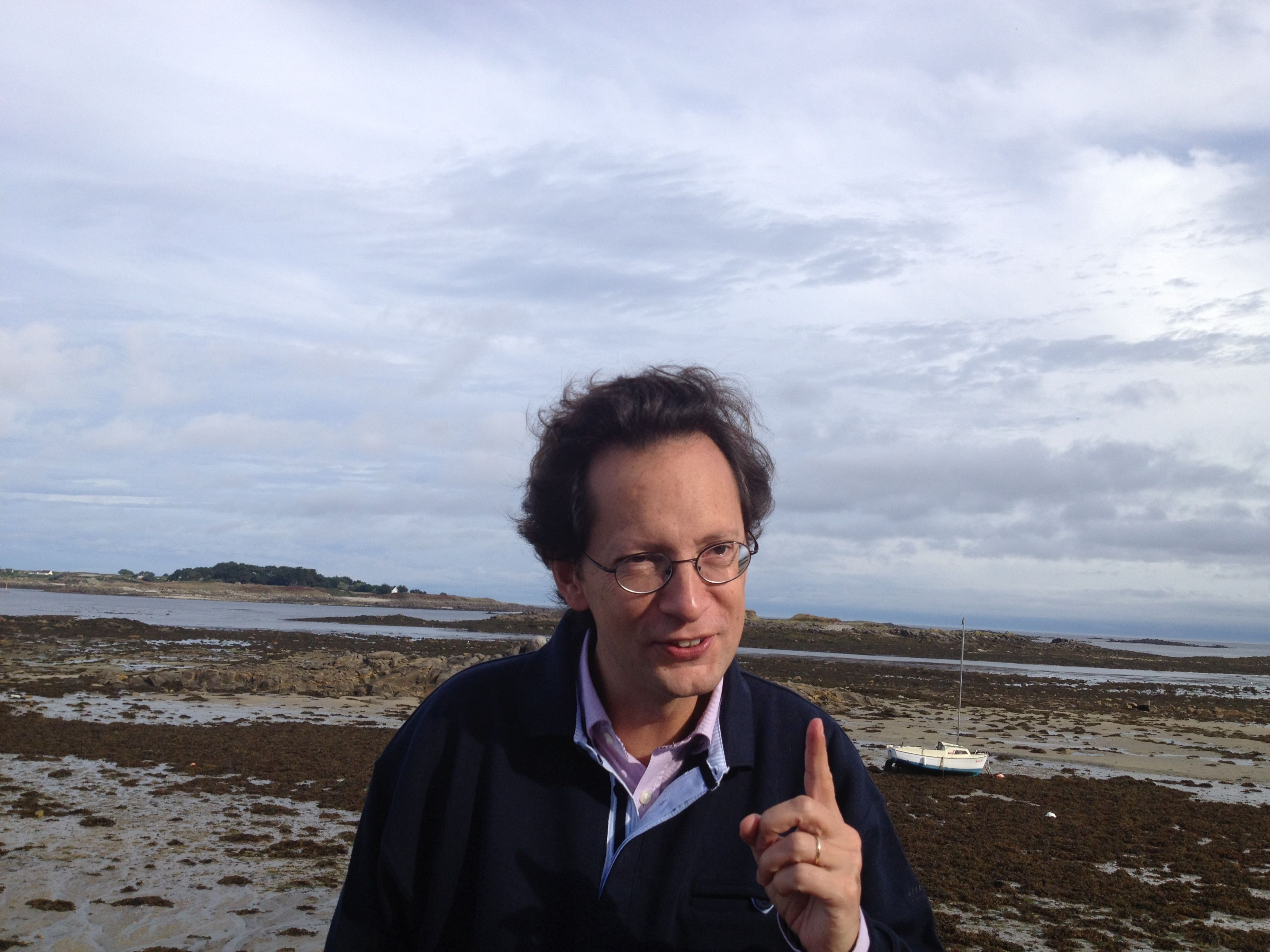}
\end{center}

\end{document}